\documentclass[iop,tighten,apj]{emulateapj}

\slugcomment{Submitted to ApJ}
\shorttitle{Cooling of the diffuse ISM}
\shortauthors{Roy et al.}

\begin{document}

\title{C~{\sc ii} radiative cooling of the Galactic diffuse interstellar medium: Insight about the star formation in Damped Lyman-$\alpha$ systems}

\author{Nirupam Roy\altaffilmark{1}, Stephan Frank\altaffilmark{2}, Christopher L. Carilli\altaffilmark{3,4}, Smita Mathur\altaffilmark{2,5}, Karl M. Menten\altaffilmark{6} and Arthur M. Wolfe\altaffilmark{7,8}}

\altaffiltext{1}{Department of Physics and Centre for Theoretical Studies, Indian Institute of Technology, Kharagpur 721302, India; Current affiliation: Department of Physics, Indian Institute of Science, Bangalore 560012, India; Contact author's electronic address: nroy@physics.iisc.ernet.in}
\altaffiltext{2}{Department of Astronomy, The Ohio State University, Columbus, OH 43210, USA}
\altaffiltext{3}{National Radio Astronomy Observatory, PO Box O, Socorro, NM 87801, USA}
\altaffiltext{4}{Astrophysics Group, Cavendish Laboratory, JJ Thomson Avenue, Cambridge CB3 0HE, UK}
\altaffiltext{5}{Center for Cosmology and Astro-Particle Physics, The Ohio State University, Columbus, OH 43210, USA}
\altaffiltext{6}{Max-Planck-Institut f\"{u}r Radioastronomie, Auf dem H\"{u}gel 69, D-53121 Bonn, Germany}
\altaffiltext{7}{Department of Physics and Center for Astrophysics and Space Sciences, University of California, San Diego, La Jolla, CA 92093, USA}
\altaffiltext{8}{Deceased 2014 February 17.}

\begin{abstract}
The far-infrared [C~{\sc ii}] 158 $\mu$m fine structure transition is 
considered to be a dominant coolant in the interstellar medium. For this 
reason, under the assumption of a thermal steady state, it may be used to 
infer the heating rate and, in turn, the star formation rate in local, as well 
as in high redshift systems. In this work, radio and ultraviolet observations 
of the Galactic interstellar medium are used to understand whether C~{\sc ii} 
is indeed a good tracer of the star formation rate. For a sample of high 
Galactic latitude sightlines, direct measurements of the temperature indicate 
the presence of C~{\sc ii} in both the cold and the warm phases of the diffuse 
interstellar gas. The cold gas fraction ($\sim 10 - 50\%$ of the total neutral 
gas column density) is not negligible even at high Galactic latitude. It is 
shown that, to correctly estimate the star formation rate, C~{\sc ii} cooling 
in both the phases should hence be considered. The simple assumption, that the 
[C~{\sc ii}] line originates only from either the cold {\it or} the warm phase, 
significantly underpredicts or overpredicts the star formation rate, 
respectively. These results are particularly important in the context of the 
Damped Lyman-$\alpha$ systems for which a similar method is often used to 
estimate the star formation rate. The derived star formation rates in such 
cases may not be reliable if the temperature of the gas under consideration is 
not constrained independently.
\end{abstract}


\section{Introduction}
\label{sec:int}

In the standard model of the Galactic diffuse interstellar medium (ISM), a 
balance of the heating and cooling processes leads to a thermal steady state 
\citep*[e.g.][]{fi65,fi69,mo77,wo95,wo03}. Thus, in a multiphase medium, 
different phases coexist at different temperature but in an approximate thermal 
pressure equilibrium. Broadly, the diffuse multiphase medium consists of the 
cold neutral medium (CNM), the warm neutral medium (WNM), the warm ionized 
medium (WIM) and the hot ionized medium (HIM). The local physical conditions 
of the ISM are determined by a host of factors, including the local radiation 
field and cosmic ray energy density, the dust grain abundance, composition and 
size distribution, material and mechanical energy transfer from both impulsive 
disturbances such as the ejection of the outer mantle in the late stages of 
stellar evolution and supernova explosions, as well as more steady sources 
such as stellar winds. 

Here we focus on understanding the [C~{\sc ii}] 158 $\mu$m fine structure 
cooling in the diffuse ISM. Based on the assumption of thermal steady state, 
observation of C~{\sc ii}$^*$ absorption can be used to estimate the 
cooling/heating rate and, in turn, to infer the star formation rate (SFR). In 
detail however, for a given estimated cooling rate, the inferred ultraviolet 
(UV) and cosmic ray flux (and thus, the SFR) change significantly depending on 
the assumed physical conditions of the gas. Whether the gas is in the cold or 
warm phase, for example, is one of the important factors in this regard. Thus, 
to estimate the SFR from C~{\sc ii}$^*$ absorption, direct measurement of the 
temperature of the diffuse ISM for the same lines of sight is also necessary.

In this work we present temperature measurements from Galactic H~{\sc i} 21 cm 
observations toward a sample of high latitude extra-galactic radio sources. 
For these lines of sight, both H~{\sc i} 21 cm single dish emission spectra 
and UV spectroscopic data covering C~{\sc ii$^*$} absorption are available. 
Background motivation of this work is presented in \S\ref{sec:theory}. The 
details of observation and analysis techniques used here are described in 
\S\ref{sec:obs}. The results and relevant discussions are presented in 
\S\ref{sec:res} and \S\ref{sec:dis}, respectively. Finally, we summarize the 
conclusions in \S\ref{sec:con}.

\section{Background}
\label{sec:theory}

\begin{figure}
\begin{center}
\includegraphics[scale=0.33, angle=-90.0]{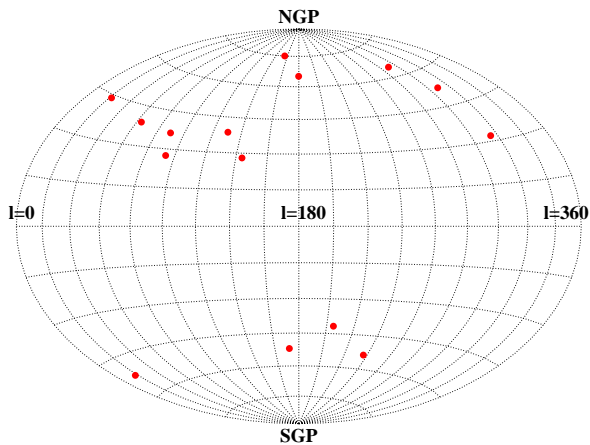}
\caption{\label{fig:fig1} Sky distribution of the observed lines of sight shown in the Galactic coordinate system with Hammer projection.}
\end{center}
\end{figure}

\subsection{Heating and cooling of the diffuse ISM}

Major sources of heating in the diffuse ISM are (i) photoelectric heating due 
to ejection of electrons from the dust grains by the far ultraviolet (FUV) 
radiation field, (ii) heating due to ionization by cosmic rays and soft X-rays, 
(iii) photoionization of species like C~{\sc i}, Si~{\sc i}, Fe~{\sc i} 
etc.~(for which the ionization potential is less than 13.6 eV) by the FUV 
radiation field and (iv) collisional ionization of H and He by impact with H 
and $e^-$ (e.g., \citealt{fi69,hm89,bt94,wo95}; \citealt*{wol03}). Since both 
photoelectric heating and heating due to ionization by cosmic rays and X-rays 
are related to the interstellar radiation field, the total heating rate is a 
function of the SFR ($\dot{\psi_*})$. Dominant cooling mechanisms in the ISM, 
on the other hand, are (i) cooling by the fine structure lines of C~{\sc i}, 
C~{\sc ii}, O~{\sc i}, Si~{\sc i}, Si~{\sc ii}, S~{\sc i}, Fe~{\sc i} and 
Fe~{\sc ii}, (ii) metastable transitions of C~{\sc ii}, O~{\sc i}, Si~{\sc ii}, 
S~{\sc ii} etc.~(iii) collisional excitation of Lyman-$\alpha$ and (iv) 
radiative recombination of $e^-$ onto dust grains and polycyclic aromatic 
hydrocarbons (PAHs) \citep[e.g.,][and references therein]{bt94,wo95,wol03}. 
Various cooling mechanisms become important at different temperature and thus 
the total cooling rate depends on the physical conditions like temperature and 
density of the gas. In thermal steady state, the total cooling rate is equal 
to the total heating rate, and the physical conditions in different phases can 
be deduced by considering the thermal and ionization equilibrium.

\subsection{Cooling by the [C{\sc ii}] fine structure line}

The [C{\sc ii}] 158 $\mu$m transition is a dominant contributor to the cooling 
in the ISM because of (1) the high abundance of carbon (second most abundant 
metal in gas phase), (2) high abundance of its singly ionized stage, (3) 
relatively low optical depth of the transition, and (4) the easy excitation of 
the $^2P_{3/2}$ fine structure state ($h\nu/k = 91$ K) by collisions under 
typical conditions in the diffuse ISM. Please see \citet{gold12} and references 
therein for an extensive review.

The C~{\sc ii} radiative cooling rate can be determined directly from the 
[C~{\sc ii}] line intensity of the $^2P_{3/2}$ to $^2P_{1/2}$ 157.7 $\mu$m 
transition in the far-infrared (FIR). Alternately, the measured column 
density of C~{\sc ii}$^*$ per H~{\sc i} atom is also believed to be a direct 
measure of the cooling rate of the gas (\citealt*{pwd79}; \citealt{wol03}; 
\citealt*{le04}). C~{\sc ii}$^*$ column density can be measured from the 
C~{\sc ii}$^*$ absorption lines at 1037.018 \AA~ and 1335.708 \AA~ in the FUV 
originating in the same $^2P_{3/2}$ state. 

\subsection{C~{\sc ii}: An estimator of the star formation rate}

For thermal steady state condition, one can infer the heating rate from the 
measured C{\sc ii} cooling rate. The heating rate, in turn, depends on the 
fluxes of UV photons and cosmic rays, and thus on the SFR. Based on this 
reasoning, observation of C~{\sc ii} or C~{\sc ii}$^*$ is often considered as 
a tracer of star formation in the local as well as in the high redshift 
Universe \citep[e.g.][]{bose02,rodr06,stac10,kapa14,pine14,sarg12,sarg14}. A 
similar argument has been used for high redshift damped Lyman-$\alpha$ systems 
(DLAs) to deduce $\dot{\psi_*}$ \citep[e.g.][]{wol03,wol04}.

One of the main sources of uncertainty, while estimating the SFR for DLAs, is 
the physical condition, more importantly the temperature, of the gas. For 
example, in absence of any direct measurement of temperature, the SFR 
calculations for the DLAs by \citet{wol03} were based on the assumption that 
most of the gas giving rise to the C~{\sc ii}$^*$ absorption is in the CNM 
phase. Otherwise, if it is assumed that all the gas is in the WNM phase, then 
the inferred SFR per unit area in DLAs is significantly higher than that of 
the Milky way. On the other hand, \citet{le04} studied the C~{\sc ii}$^*$ 
absorption along high Galactic latitude extra-galactic sources, and, in 
conjunction with observations of the diffuse H$\alpha$ emission along these 
same line of sight, concluded that most of the C~{\sc ii}$^*$ absorption 
occurs in the WNM or the WIM. This conclusion, that most of the C~{\sc ii}$^*$ 
along high Galactic latitudes comes from the WNM/WIM, rests in large part on 
the assumption that there is negligible amount of gas in the CNM phase along 
these sight lines. However, there exist several high latitude lines of sight 
with a high CNM fraction \citep[e.g.][and references therein]{roy06,roy13a,roy13b}. 

One way to critically re-examine this uncertainty regarding the inferred SFR, 
would be to directly measure the temperature of the diffuse H~{\sc i} in our 
Galaxy for lines of sight with C~{\sc ii}$^*$ absorption. This will constrain 
the cold gas fraction for lines of sight with H~{\sc i} column density similar 
to that of DLAs. With this, and the known C~{\sc ii}$^*$ column density, it 
will also be possible, following the same reasoning used for the DLAs, to 
estimate the SFR, and compare it with the Galactic SFR derived using other 
methods.

\section{Summary of the data and the analysis}
\label{sec:obs}

\begin{table*}
 \caption{Details of the sample}
\begin{center}
 \begin{tabular}{lllllll}
 \tableline
Background & \multicolumn{1}{c}{Coordinate} & E(B-V) & N(C{\sc ii}$^*$) & N(H$_2$) & N(H~{\sc i}) \\
Sources    & \multicolumn{1}{c}{l, b [deg]} & [mag]  & \multicolumn{3}{c}{[$\log$(cm$^{2}$)]} \\
\tableline    
3C~71            & 172.10, ~$-$51.93 & 0.030 &  14.02 & 18.07       & 20.43  \\
3C~273           & 289.95, ~$+$64.36 & 0.018 &  14.03 & 15.92       & 20.21  \\
3C~351           & ~90.08, ~$+$36.38 & 0.021 &  14.28 &  ---        & 20.23  \\
H~1821$+$643     & ~94.00, ~$+$27.42 & 0.038 &  14.21 & 15.99       & 20.54  \\
HB89~0716$+$714  & 143.98, ~$+$28.02 & 0.027 &  14.09 & 17.04       & 20.49  \\
HB89~1553$+$113  & ~21.91, ~$+$43.96 & 0.046 &  14.24 & 17.58       & 20.57  \\
HB89~2155$-$304  & ~17.73, ~$-$52.25 & 0.019 &  13.67 & 14.42       & 20.15  \\
Mrk~421          & 179.83, ~$+$65.03 & 0.013 &  13.97 & 14.83       & 20.18  \\
Mrk~501          & ~63.60, ~$+$38.86 & 0.017 &  $\leq$13.47 & 15.49 & 20.15  \\
NGC~1399         & 236.72, ~$-$53.64 & 0.012 &  13.76 & $\geq$14.55 & 20.14  \\
NGC~4151         & 155.08, ~$+$75.06 & 0.024 &  14.16 & 16.60       & 20.40  \\
NGC~5236         & 314.58, ~$+$31.97 & 0.058 &  15.00 & 14.79       & 20.56  \\
PG~1100$+$772    & 130.39, ~$+$38.55 & 0.030 &  $\leq$13.47 & 19.09 & 20.43  \\
PG~1302$-$102    & 308.59, ~$+$52.16 & 0.037 &  14.00 & 16.30       & 20.51  \\
PKS~0405$-$12    & 204.93, ~$-$41.76 & 0.051 &  14.39 & 16.01       & 20.54  \\
\tableline
\end{tabular}\\
\end{center}
\label{table:table1}
\end{table*}

From the sample of \citet{le04} and \citet*{wa06} with UV spectroscopic 
observations, 15 sources were selected in the declination range accessible to 
the Giant Metrewave Radio Telescope \citep*[GMRT;][]{sw91} and the Karl G. 
Jansky Very Large Array \citep*[VLA;][]{pe11}. Figure~\ref{fig:fig1} shows the 
position of the background sources in the Galactic coordinate system. These are 
all high Galactic latitude sources ($|b|>20^\circ$), with 1.4 GHz flux density 
greater than 100 mJy and a substantial flux in the compact components. 
Table~\ref{table:table1} lists the names of the background sources, Galactic 
coordinates, interstellar reddening, and the column density of H~{\sc i}, 
H$_2$ and C~{\sc ii}$^*$ for these lines of sight. High spectral resolution 
H~{\sc i} emission spectra along these lines of sight are already available 
from the Leiden/Argentine/Bonn (LAB) Galactic H~{\sc i} survey \citep*{hb97,ar00,ba05,ka05}. The H~{\sc i} column densities from the LAB survey for these lines 
of sight are $(1.4 - 3.7)\times10^{20} {\rm ~cm}^{-2}$. So, for extra-galactic 
sources, similar lines of sight will have $\sim (3 - 7)\times10^{20} {\rm ~cm}^{-2}$ H~{\sc i} column density - very much like the typical DLA lines of sight.

\begin{figure}
\begin{center}
\includegraphics[scale=0.33, angle=-90.0]{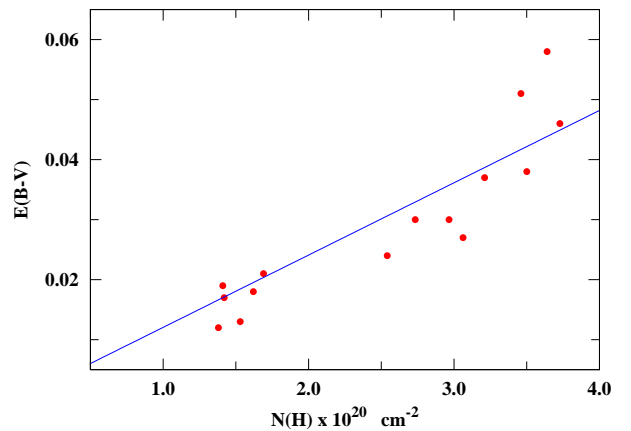}
\caption{\label{fig:fig2} Reddening with total hydrogen column density for the lines of sight. Solid line is expected $E(B-V)/N(H)$ ratio from \citet{lh14}.}
\end{center}
\end{figure}

Figure~\ref{fig:fig2} shows the relation between total hydrogen column density 
and reddening for the lines of sight in our sample. The Galactic reddening 
E(B-V) is derived from the infrared dust maps with recent recalibration \citep*{sfd98,sf11}. The total hydrogen column density N(H) = N(H~{\sc i}) + 2N(H$_2$) 
is based on LAB N(H~{\sc i}), and N(H$_2$) from \citet{wa06}. The solid line 
corresponds to the expected value of N(H~{\sc i})/E(B-V) = $8.3\times10^{21} 
{\rm ~cm}^{-2} {\rm mag}^{-1}$ \citep{lh14}. This gas column density to 
reddening ratio is $\sim 40$\% higher than the corresponding value derived, 
e.g., by \citet*{boh78}, from optical/UV observations. However, for this high 
latitude sample, it is more appropriate to instead compare with the value 
derived from the radio/IR observations for similar lines of sight by 
\citet{lh14}. From Figure~\ref{fig:fig2}, reddening for these lines of sight 
seems to have no unusual deviation from the typical dust to gas ratio.

\subsection{Radio observations and the data reduction}

\begin{figure*}
\begin{center}
\includegraphics[scale=0.45, angle=-90.0]{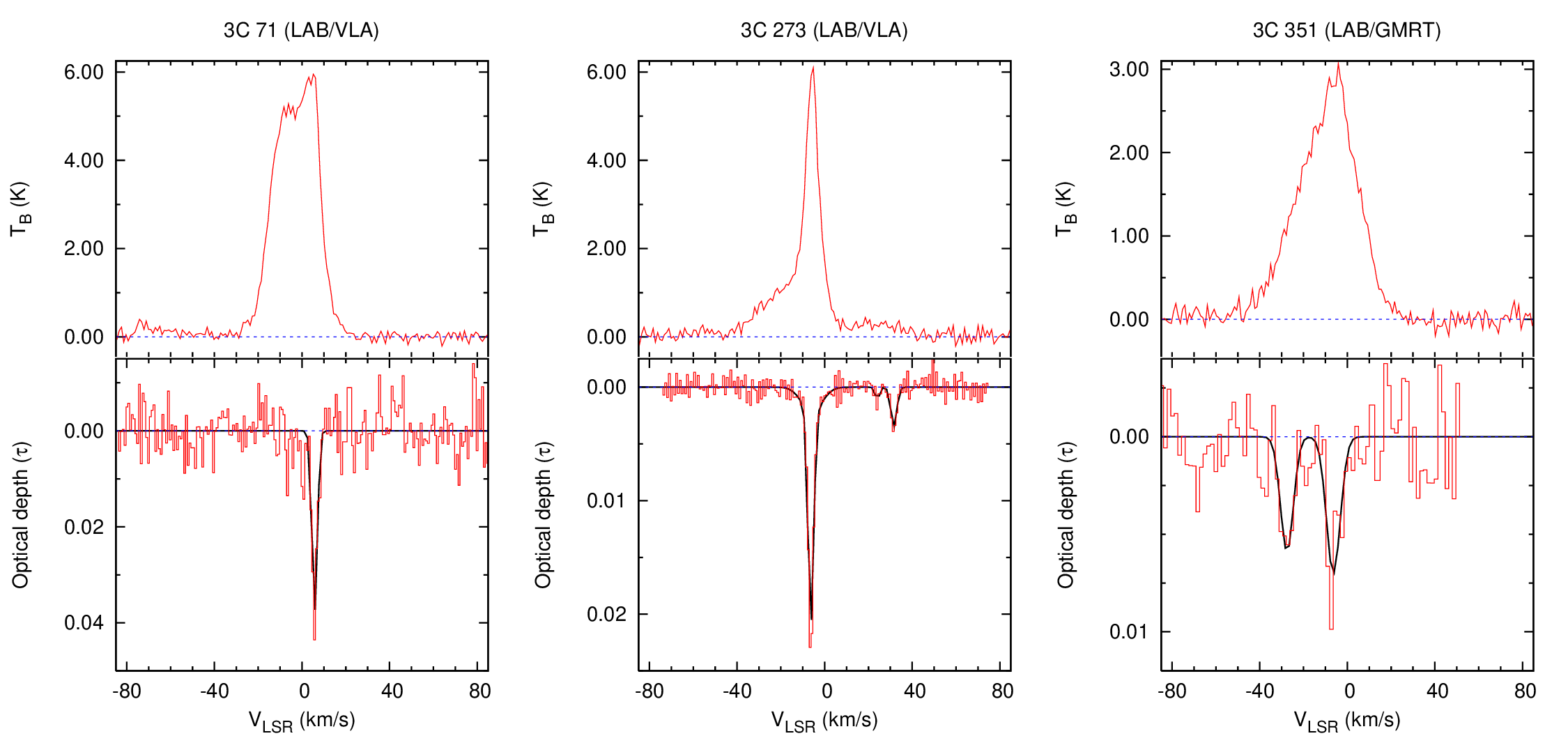}
\includegraphics[scale=0.45, angle=-90.0]{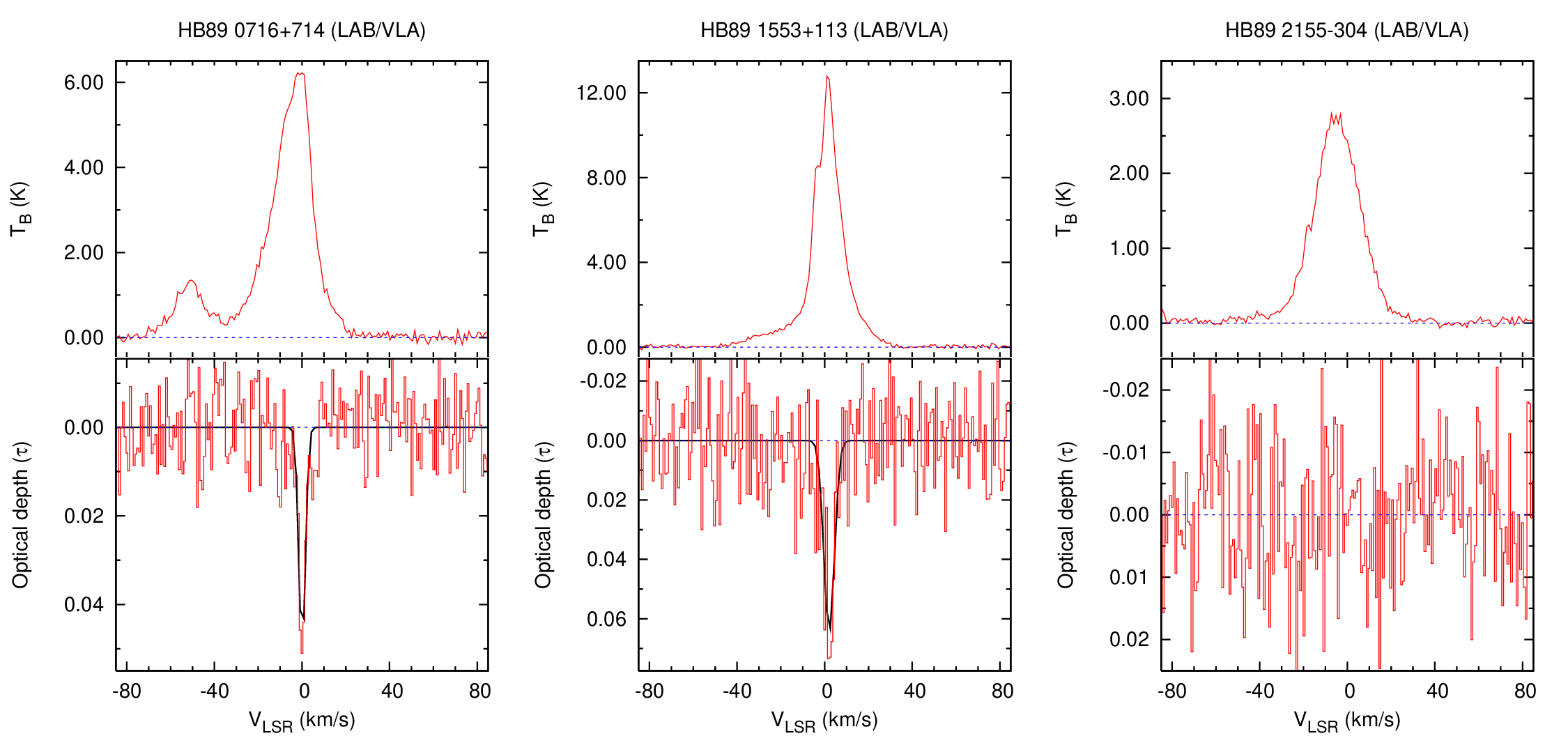}
\caption{\label{fig:fig3} H~{\sc i} emission/absorption spectra for the sample. In each case, the top panel shows emission spectrum from the LAB survey, and the bottom panel shows absorption spectrum from the GMRT/VLA data. Velocity resolution is $\sim1.0$, $0.8$ and $1.6$ km~s$^{-1}$ for LAB, VLA and GMRT spectra, respectively.}
\end{center}
\end{figure*}

\setcounter{figure}{2}
\begin{figure*}
\begin{center}
\vspace*{-2mm}\includegraphics[scale=0.45, angle=-90.0]{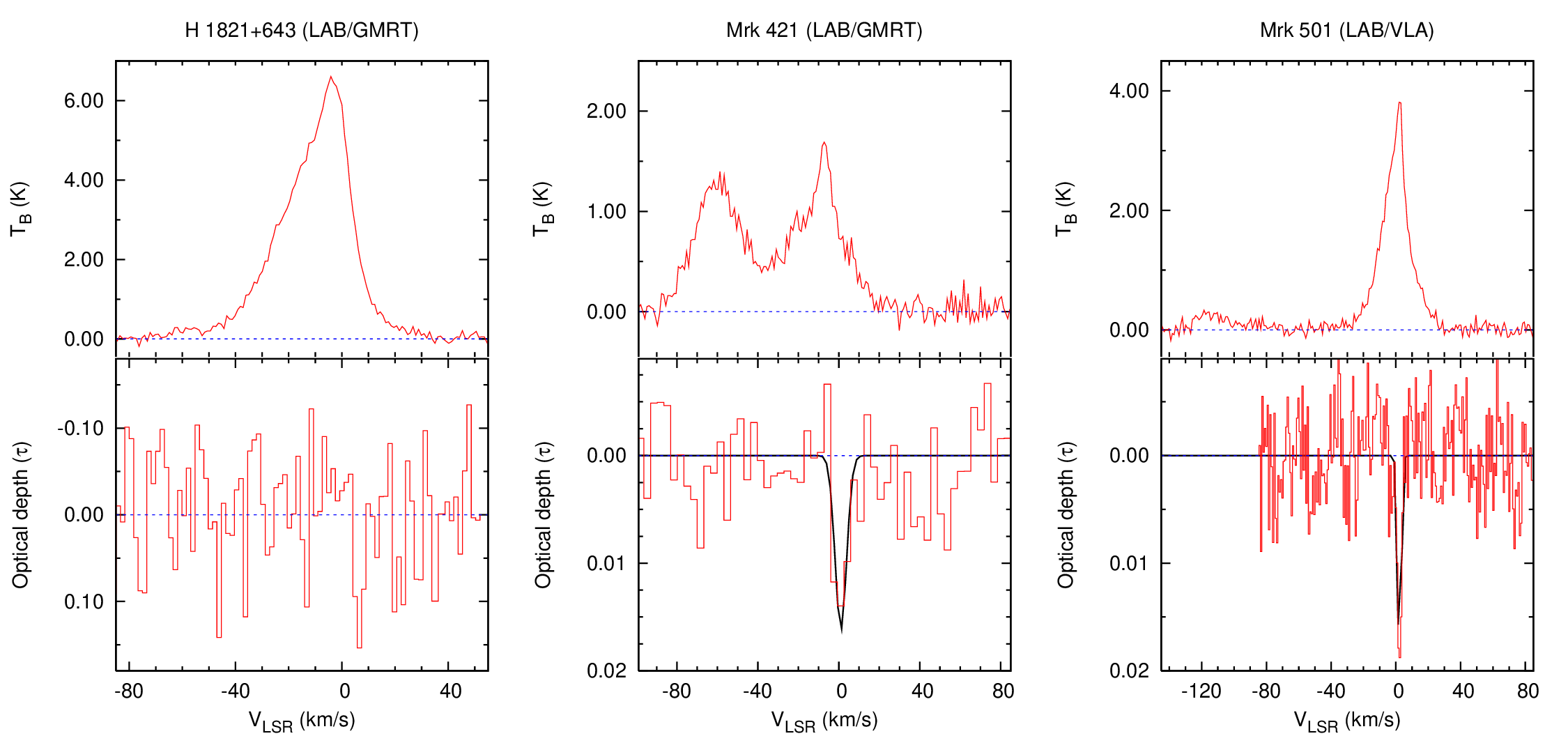}
\includegraphics[scale=0.45, angle=-90.0]{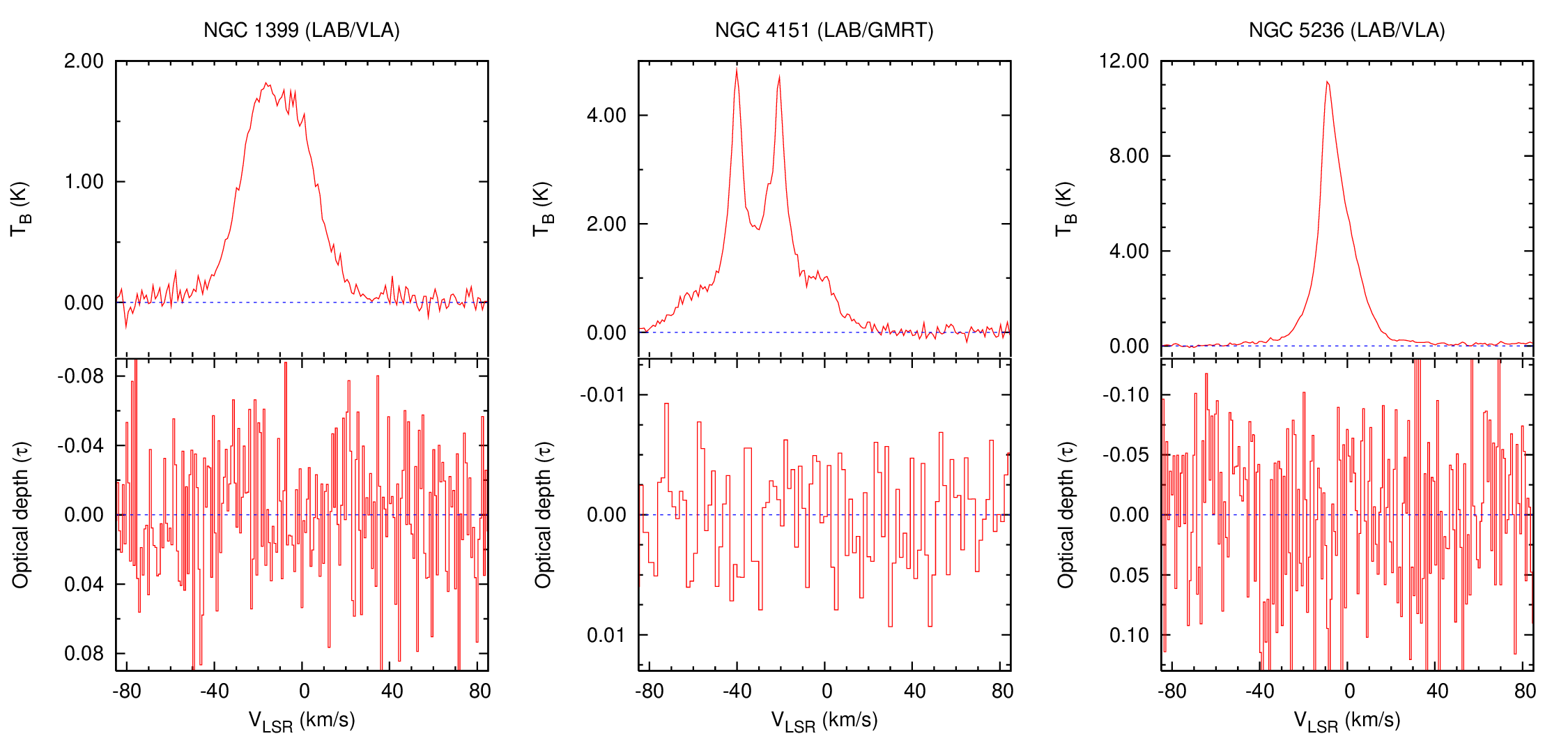}
\includegraphics[scale=0.45, angle=-90.0]{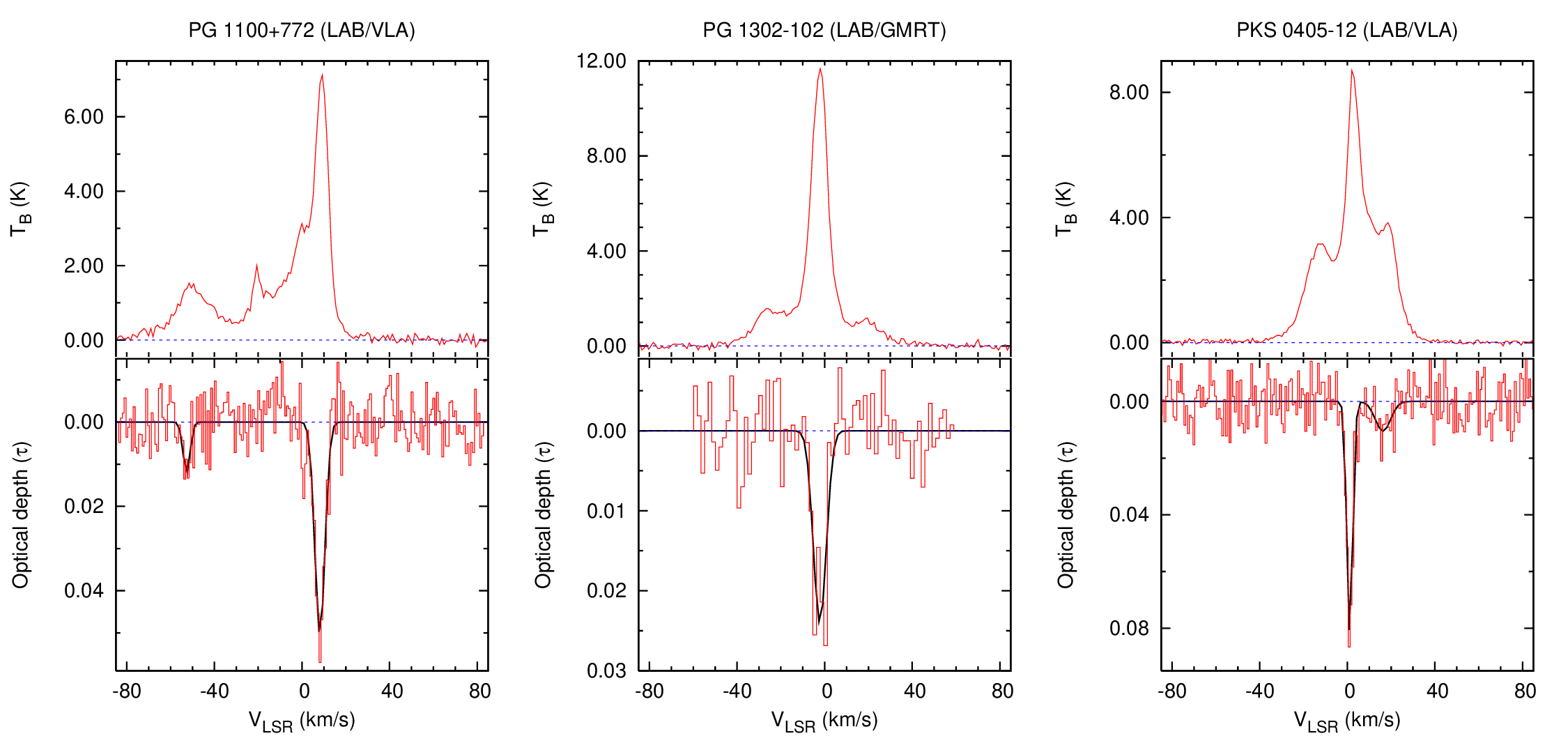}
\caption{\label{fig:fig3a} H~{\sc i} emission and absorption spectra for the sample (continued).}
\end{center}
\end{figure*}

The GMRT observations were carried out in cycle 8 (2005) for five of the 
sources using a total 1.0~MHz bandwidth with 128 spectral channels (i.e. a 
velocity resolution of $\sim$ 1.6 km~s$^{-1}$). The VLA B-configuration 
observations (project code 12A-428) for ten sources were carried out in 2012 
using 256 channels over 1.0~MHz bandwidth (i.e. $\sim$ 0.8 km~s$^{-1}$ per 
channel). Depending on the target continuum flux density, on-source time was 
from 20 minutes to 6 hours. Short scans on calibrator sources were used for 
flux calibration, phase calibration and also to determine the bandpass shape. 
Unfortunately, a significant amount of data are affected by interference, and 
has to be excluded. Standard data analysis including flagging bad data, 
calibration, and imaging was done using the Astronomical Image Processing 
System ({\small AIPS}; produced and maintained by the National Radio Astronomy 
Observatory). The continuum emission, estimated by averaging data from 
line-free channels, was subtracted from the multi-channel visibility data. The 
residual data were then used to make the image cubes, and any small residual 
continuum was subtracted in the image plane by fitting a linear baseline to 
the line-free regions. The absorption spectra toward the compact component 
were then extracted from the high resolution image cubes where the smooth 
H~{\sc i} emission was resolved out. Finally, the absorption spectra were 
converted from flux density to optical depth ($\tau$) using the 1.4 GHz flux 
density value at the corresponding location of the continuum image.

The H~{\sc i} emission and absorption spectra for all 15 lines of sight are 
shown in Figure~\ref{fig:fig3}. For each line of sight, the top and the bottom 
panels show the LAB H~{\sc i} emission spectra and the GMRT or VLA H~{\sc i} 
absorption spectra, respectively. Name of the background continuum source and 
the telescope names are also mentioned at the top. For 10 out of these 15 lines 
of sight, H~{\sc i} 21 cm absorption is clearly detected. For the detections, 
(multi-)Gaussian model of the absorption spectra are also overplotted in 
Figure~\ref{fig:fig3}.

\subsection{Temperature estimation}

In the radio regime, the classical method to determine the temperature of the 
gas consists of observing the H~{\sc i} 21 cm line in absorption towards a 
bright radio continuum source, and 21 cm emission spectrum along a nearby line 
of sight. H~{\sc i} emission and absorption spectra allow one to measure the 
spin temperature \citep*[T$_{\rm s}$; see e.g.][for details]{kh88}, which is 
often used as a proxy for the kinetic temperature (T$_{\rm k}$) of the gas. 
For the CNM, T$_{\rm s}$ is expected to be tightly coupled to T$_{\rm k}$ via 
collisions \citep*{fi58}.

It is also possible to estimate the temperature from the observed linewidth of 
the H~{\sc i} emission and absorption components. This method is more useful 
for absorption spectra where the opacity is additive for a multi-Gaussian 
component fit. For emission spectra, the relative position of different 
components along the line of sight being a-priori unknown, multi-component 
decomposition is more complicated, and often there is no straightforward and 
unique interpretation. Due to possible non-thermal broadening, the observed 
linewidth provides only an upper limit to T$_{\rm k}$.

For the 10 cases with a detection in the present sample, Gaussian components 
are fitted to the absorption spectra. Table~\ref{table:table11} presents the 
integrated H~{\sc i} column densities (from the LAB survey), the integrated 
optical depth values (or upper limits) from this study, and also shows the 
best fit parameter values (peak optical depth, centre and width of the 
components) for all the spectra. The width of each component then provides an 
upper limit T$_{\rm k, max}$. We also used the emission and the absorption 
spectra to compute spin temperature spectra (at a resolution of $\sim 1.0$ and 
1.6 km~s$^{-1}$ for VLA and GMRT sample, respectively). The local minimum of 
spin temperature, T$_{\rm s, min}$ from these spectra over the velocity range 
of any absorption component is taken as an estimator of T$_{\rm s}$ for the 
corresponding ``cold'' component. For velocity intervals with only H~{\sc i} 
emission (and corresponding C~{\sc ii}$^*$ absorption; see below), but no 
detections of H~{\sc i} absorption, the same method is used to estimate the 
lower limit of T$_{\rm s}$ from the $3\sigma$ upper limit of the optical depth. Please note that the H~{\sc i} emission may have some contribution due to 
blending with components unrelated to the absorption. Hence, in general, using 
T$_{\rm s, min}$ as an estimator of T$_{\rm s}$ may result in an overestimation 
of temperature. Thus, low T$_{\rm s}$ values will be a conservative indicator 
of cold gas.

\begin{table*}
 \caption{Details of the absorption spectra and the Gaussian fit parameters}
\begin{center}
 \begin{tabular}{lccccc}
 \tableline
Background &  N(H~{\sc i})$^a$ & $\int\tau dv$~$^b$ & \multicolumn{3}{c}{Best fit parameters}  \\
Sources    &  $10^{19}$ cm$^{2}$ & km~s$^{-1}$   &  $\tau_{\rm peak}$ & $v_c$ (km~s$^{-1}$) & $\sigma_v$ (km~s$^{-1}$) \\
\tableline    
3C~71                & 27.1 & 0.120 $\pm$ 0.032 & 0.0380 $\pm$ 0.0041 &  5.76 $\pm$ 0.15 & 1.26 $\pm$ 0.15 \\
3C~273               & 16.2 & 0.103 $\pm$ 0.008 & 0.0022 $\pm$ 0.0002 & -5.86 $\pm$ 0.23 & 4.64 $\pm$ 0.32 \\
                  &  &      & 0.0187 $\pm$ 0.0003 & -6.28 $\pm$ 0.01 & 1.32 $\pm$ 0.02 \\
                  &  &      & 0.0035 $\pm$ 0.0002 & 31.59 $\pm$ 0.07 & 1.36 $\pm$ 0.07 \\
                  &  &      & 0.0011 $\pm$ 0.0002 & 24.07 $\pm$ 0.19 & 0.98 $\pm$ 0.19 \\
3C~351               & 16.9 & 0.100 $\pm$ 0.015 & 0.0070 $\pm$ 0.0012 & -6.36 $\pm$ 0.65 & 3.19 $\pm$ 0.65 \\
                  &  &      & 0.0059 $\pm$ 0.0013 &-27.57 $\pm$ 0.73 & 2.95 $\pm$ 0.73 \\
H~1821$+$643         & 35.0 & $\lesssim$  0.172 & \multicolumn{3}{c}{ }                                    \\
HB89~0716$+$714      & 30.6 & 0.189 $\pm$ 0.088 & 0.0496 $\pm$ 0.0054 &  0.06 $\pm$ 0.19 & 1.52 $\pm$ 0.19 \\
HB89~1553$+$113      & 37.2 & 0.389 $\pm$ 0.081 & 0.0646 $\pm$ 0.0089 &  2.06 $\pm$ 0.39 & 2.40 $\pm$ 0.38 \\
HB89~2155$-$304      & 14.1 & $\lesssim$  0.176 & \multicolumn{3}{c}{ }                                    \\
Mrk~421              & 15.3 & 0.115 $\pm$ 0.069 & 0.0162 $\pm$ 0.0012 &  1.01 $\pm$ 0.03 & 2.84 $\pm$ 0.24 \\
Mrk~501              & 14.2 & 0.060 $\pm$ 0.047 & 0.0201 $\pm$ 0.0031 &  2.20 $\pm$ 0.21 & 1.19 $\pm$ 0.21 \\
NGC~1399             & 13.8 & $\lesssim$  0.714 & \multicolumn{3}{c}{ }                                    \\
NGC~4151             & 25.4 & $\lesssim$  0.131 & \multicolumn{3}{c}{ }                                    \\
NGC~5236             & 36.4 & $\lesssim$  1.078 & \multicolumn{3}{c}{ }                                    \\
PG~1100$+$772        & 27.2 & 0.343 $\pm$ 0.066 & 0.0507 $\pm$ 0.0037 &  8.18 $\pm$ 0.19 & 2.31 $\pm$ 0.19 \\
                  &  &      & 0.0121 $\pm$ 0.0039 &-52.68 $\pm$ 0.60 & 1.60 $\pm$ 0.60 \\
PG~1302$-$102        & 32.1 & 0.179 $\pm$ 0.038 & 0.0240 $\pm$ 0.0029 & -2.19 $\pm$ 0.42 & 2.97 $\pm$ 0.42 \\
PKS~0405$-$12        & 34.6 & 0.374 $\pm$ 0.044 & 0.0822 $\pm$ 0.0051 &  1.13 $\pm$ 0.10 & 1.37 $\pm$ 0.10 \\
                  &  &      & 0.0105 $\pm$ 0.0032 & 16.29 $\pm$ 1.22 & 3.49 $\pm$ 1.22 \\
\tableline
\end{tabular}\\
\end{center}
\begin{flushleft}
$^a$ Total N(H~{\sc i}) with $\lesssim10\%$ uncertainty from the LAB survey;~~$^b$ Error for $\int\tau dv^{a}$ includes contribution from the full velocity range of H~{\sc i} emission.
\end{flushleft}
\label{table:table11}
\end{table*}

\begin{figure}
\begin{center}
\includegraphics[scale=0.33, angle=-90.0]{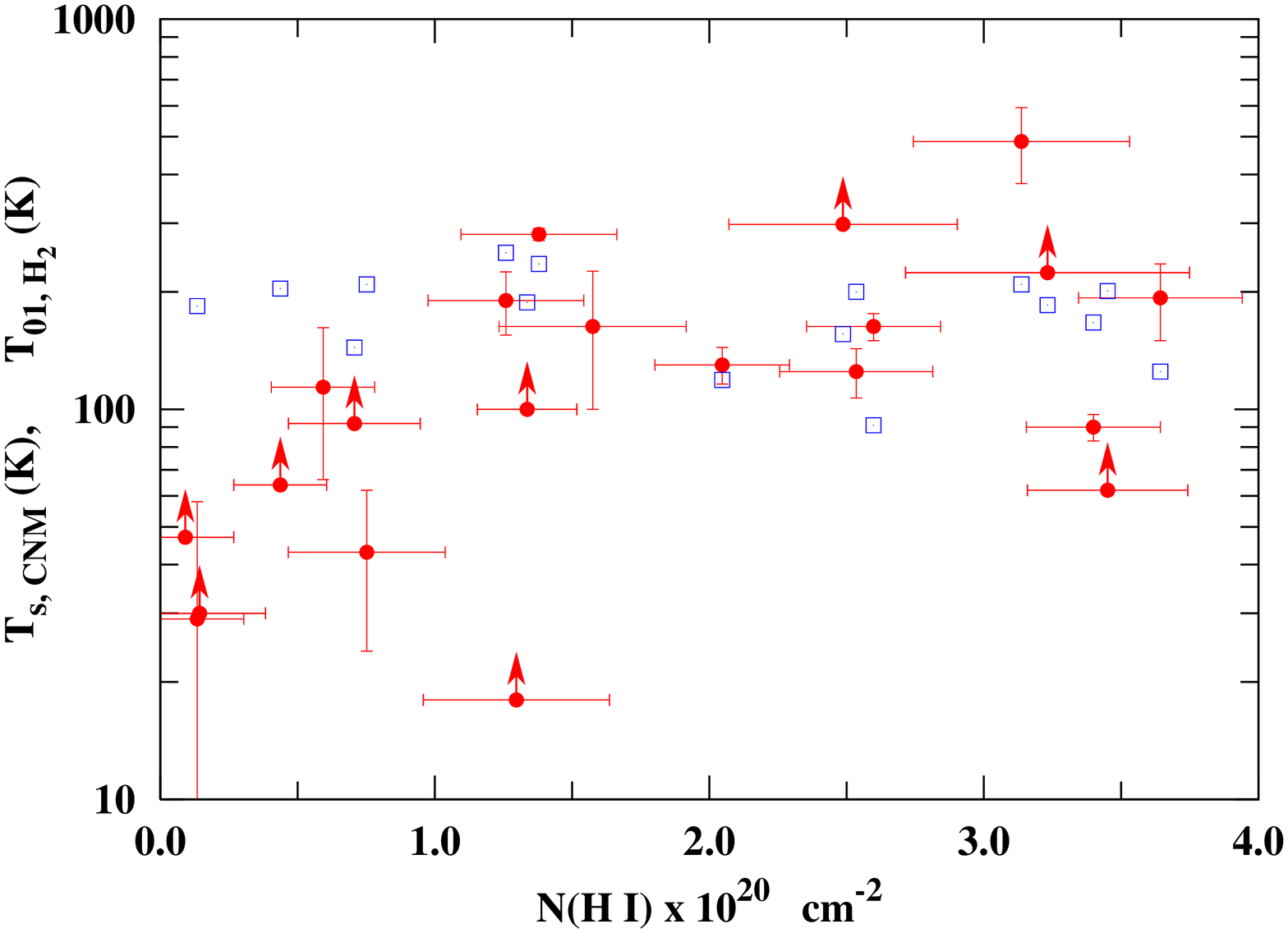}
\caption{\label{fig:fig4} Estimated H~{\sc i} spin temperature (T$_{\rm s}$) of the cold component, and the ortho-para temperature (T$_{\rm 01}$) of molecular hydrogen vs. N(H~{\sc i}) for different velocity components. T$_{\rm s}$ measurements, from this work, are shown as filled circles with errorbars (and with arrow for lower limits). T$_{\rm 01}$ from \citet{wa06} for corresponding components are shown as open squares.}
\end{center}
\end{figure}

\begin{figure}
\begin{center}
\includegraphics[scale=0.33, angle=-90.0]{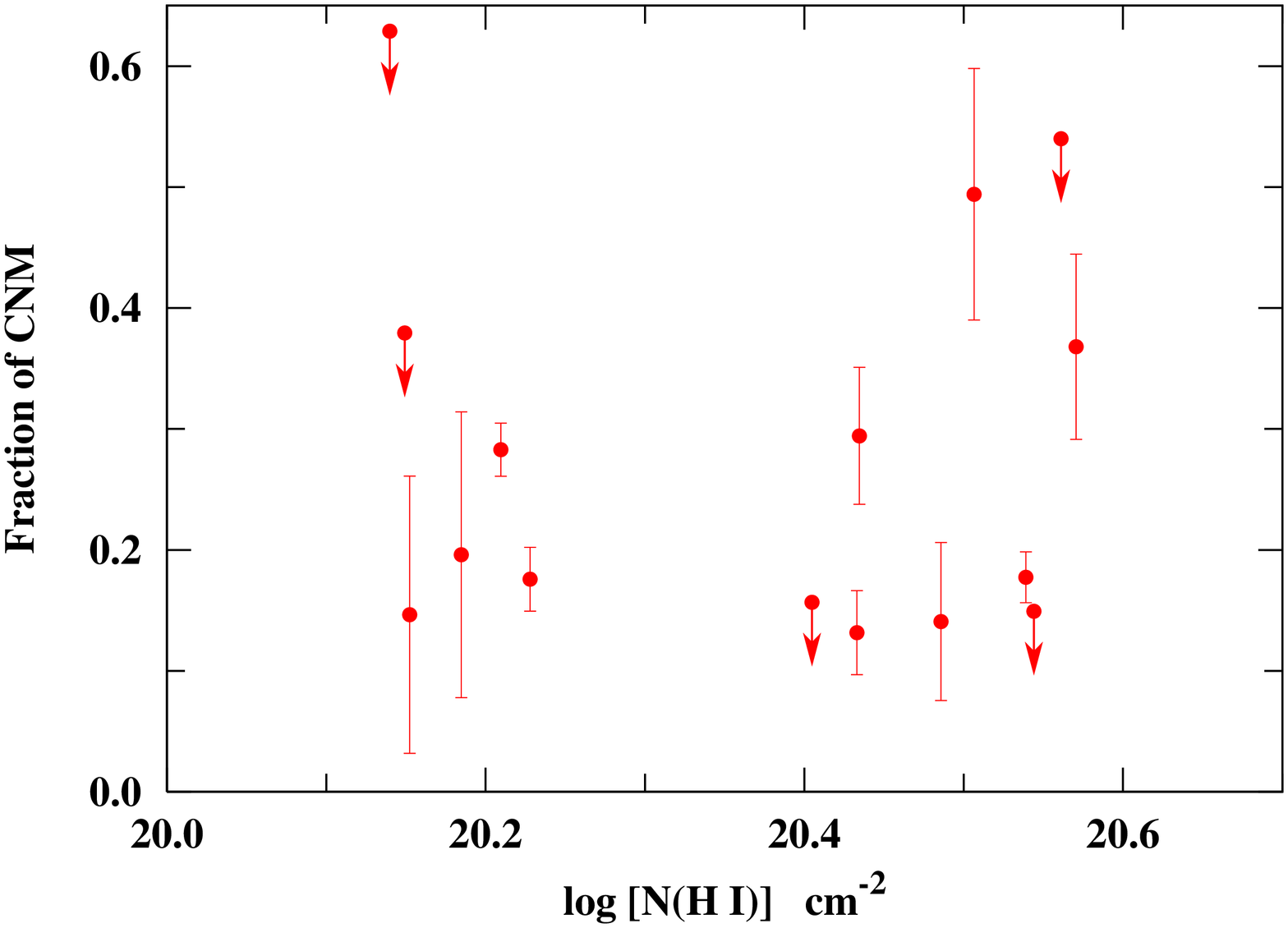}
\caption{\label{fig:fig5} Fraction of cold gas for the lines of sight estimated from H~{\sc i} absorption. For non-detection of absorption component, an average value of T$_{\rm s} = 200$ K is adopted to compute the upper limit of the cold gas column density over a velocity range of $V_{90}$ of the corresponding LAB spectra.}
\end{center}
\end{figure}

Figure~\ref{fig:fig4} presents the summary of T$_{\rm s}$ measurements for this 
sample. Considering H~{\sc i} 21 cm emission and absorption spectra, along with 
the C~{\sc ii}$^*$ absorption spectra, for these 15 lines of sight there are 
21 components with distinct velocity range, with 12 cases of detection of 
H~{\sc i} absorption, and 9 cases of non-detection. The values (and the lower 
limits) of T$_{\rm s}$, derived as outlined above, are plotted (filled circles 
with errorbars, with arrow for lower limits) against N(H~{\sc i}) for these 21 
components. Eight out of 10 components with detected H~{\sc i} absorption have 
T$_{\rm s}$ below 200 K, and only one component is above 300 K with T$_{\rm s} 
\approx 490 \pm 110$. The upper limits of T$_{\rm s}$ are not very tight due 
to low T$_{\rm B}$ and/or high RMS $\tau$. Overall, however, the presence of 
cold ISM with temperature $\lesssim 200$ K for these lines of sight is very 
clear from the T$_{\rm s}$ measurements. As expected for the turbulent ISM, 
for all the absorption components, T$_{\rm k, max}$ is always higher than 
T$_{\rm s}$ due to non-thermal broadening. 

For 16 out of these 21 components, \citet{wa06} presents measurements of 
column density, and the ``ortho-para temperature'' (T$_{\rm 01}$) of molecular 
hydrogen covering the same velocity range as the diffuse H~{\sc i}. 
T$_{\rm 01}$ is coupled to the kinetic temperature or the formation temperature 
of H$_2$ depending on whether the observed ortho-to-para ratio is achieved 
mainly by proton and hydrogen exchange collision or by reactions in the dust 
grain surface \citep*[e.g.][]{dbw73,tak01}. For the diffuse ISM with conditions 
similar to these lines of sight, T$_{\rm 01}$ is expected to trace the kinetic 
temperature \citep*{roy06}. Measured T$_{\rm 01}$ for these 16 components, 
shown in Figure~\ref{fig:fig4} as open squares, also consistently indicate the 
presence of gas with temperature around 200 K for all of these lines of sight.

\subsection{Cold and warm gas fraction}

Next, we use this estimated T$_{\rm s}$ of the detected CNM absorption 
components for each line of sight to compute the column density in the cold 
phase, and fraction of CNM (using the total H~{\sc i} column density from the 
LAB spectrum). Note that the blending effect mentioned above may cause a 
similar overestimation of CNM column density as well. Also, the total H~{\sc i} 
column density from the LAB emission spectrum is derived assuming optically 
thin condition ($\tau<<1$). For lines of sight with large optical depth, this 
may result in an underestimation of the column density \citep{chen13}. For 
these low optical depth lines of sight, however, this assumption is reasonable, 
and the correction to the total H~{\sc i} column density due to optical depth 
is negligible. For non-detection, the upper limit of the CNM column density is 
computed assuming an average T$_{\rm s} = 200$ K and a velocity width same as 
the $V_{90}$ of the corresponding LAB emission spectra. The derived CNM and WNM 
column densities are given in Table~\ref{table:table2}, and the CNM fraction 
for the sample is shown in Figure~\ref{fig:fig5}. Although in one case the cold 
gas fraction is as high as $\sim 50\%$, most of these lines of sight contains 
only $\sim 10-30\%$ CNM. In \S\ref{sec:res}, we further probe any plausible 
correlation between the C~{\sc ii}$^*$ column density and the CNM/WNM/total 
H~{\sc i} column density to understand if C~{\sc ii} cooling happens preferably 
in the cold or warm phase.

\subsection{The UV data analysis}
\label{subsec:uvda}

UV spectroscopy for these lines of sight with the {\it Far Ultraviolet 
Spectroscopic Explorer} ({\it FUSE}) and the Space Telescope Imaging 
Spectrograph (STIS) on board the {\it Hubble Space Telescope} ({\it HST}) is 
reported earlier by \citet{le04} and \citet{wa06}. However, for nine of these 
sightlines, archival data from the {\it HST} Cosmic Origin Spectrograph (COS) 
covering the C~{\sc ii}$^*$ 1335.708 \AA~ transition is now also available. 
Hence we used the COS data for these nine lines of sight, and adopted the 
column density values from literature for the remaining six lines of sight.
 
\begin{figure*}
\begin{center}
\vspace*{-2mm}\hspace*{-12mm}\includegraphics[height=165mm, angle=-90.0]{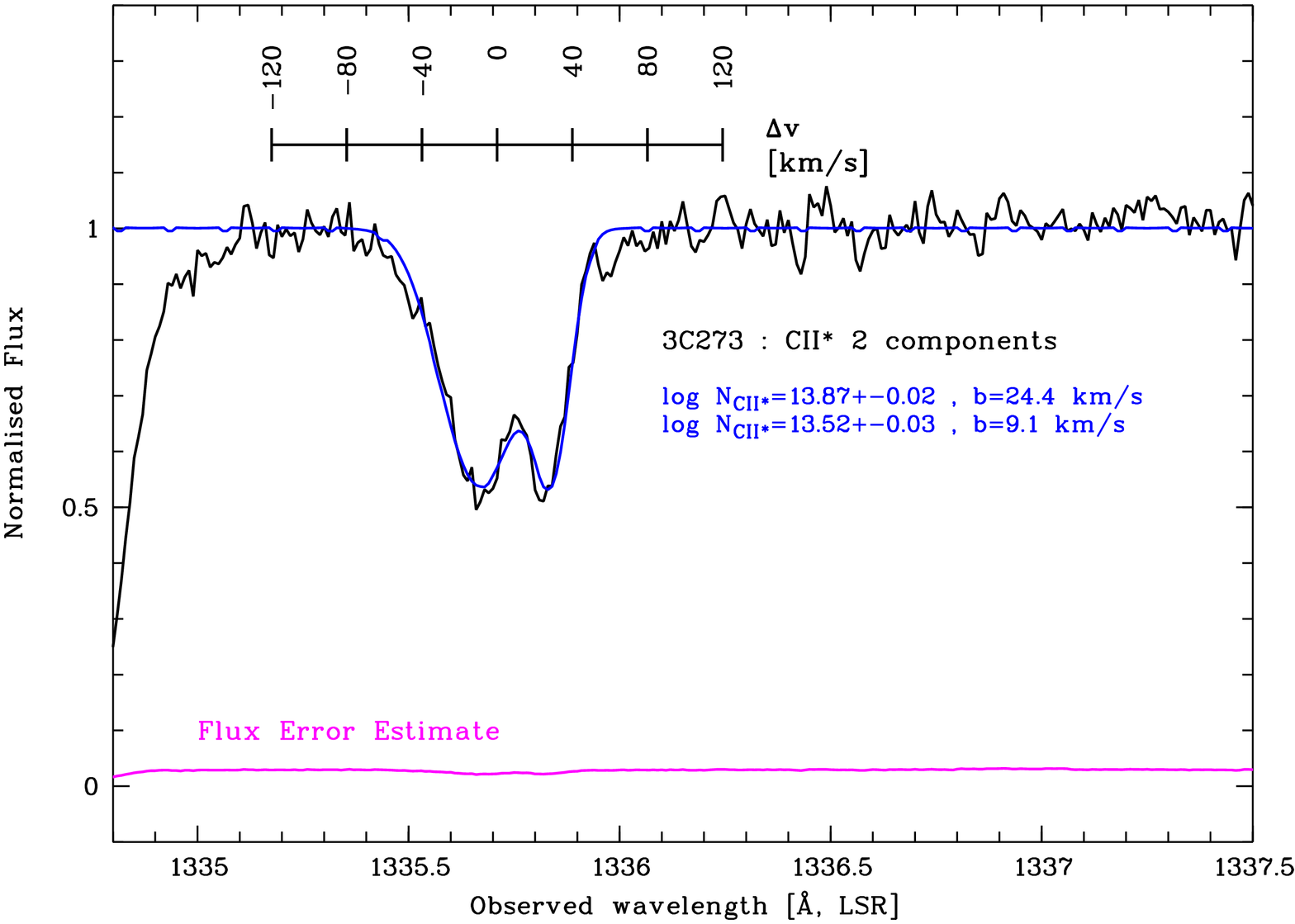}
\caption{\label{fig:exmpl} Voigt profile fitting for the Galactic C~{\sc ii}$^*$ absorption towards 3C~273. The coadded spectrum (black line) shows a velocity profile with two components (local standard-of-rest (LSR) velocity v$_{\rm LSR} \approx -8.4$ and $+29.7$ km~s$^{-1}$). The blue (magenta) line is best fit model (flux error estimate) in normalized flux unit. The corresponding LSR velocity range over the absorption profile is also shown on the plot.}
\end{center}
\end{figure*}

For the COS data, after rebinning to a common wavelength grid, and coadding 
individual exposures, we fit a local continuum to each spectrum at the 
location of the Galactic C~{\sc ii} 1334\AA~ and C~{\sc ii}$^*$ 1335\AA~ 
absorption, and then measure both the restframe equivalent width as well as 
the column density, line-centre and Doppler parameter b for the C~{\sc ii}$^*$ 
component via a Voigt profile fit. The observed spectra as well as the best fit 
models are shown in Figure~\ref{fig:exmpl}, and the total C~{\sc ii}$^*$ column 
densities from these fits are presented in Table~\ref{table:table1}. In some 
cases, certain fit parameters had to be fixed due to the complicated velocity 
structures and/or blending. Note that, for FUSE and STIS spectra of these 
sources, \citet{le04} employ two methods (equivalent width based on the curve 
of growth and apparent optical depth) very different from ours (Voigt profile 
fit), and therefore some discrepancy in the values may be expected.  

\section{Results}
\label{sec:res}

\setcounter{figure}{5}
\begin{figure*}
\begin{center}
\vspace*{-4mm}
\hspace*{-12mm}
\includegraphics[height=165mm, angle=-90.0]{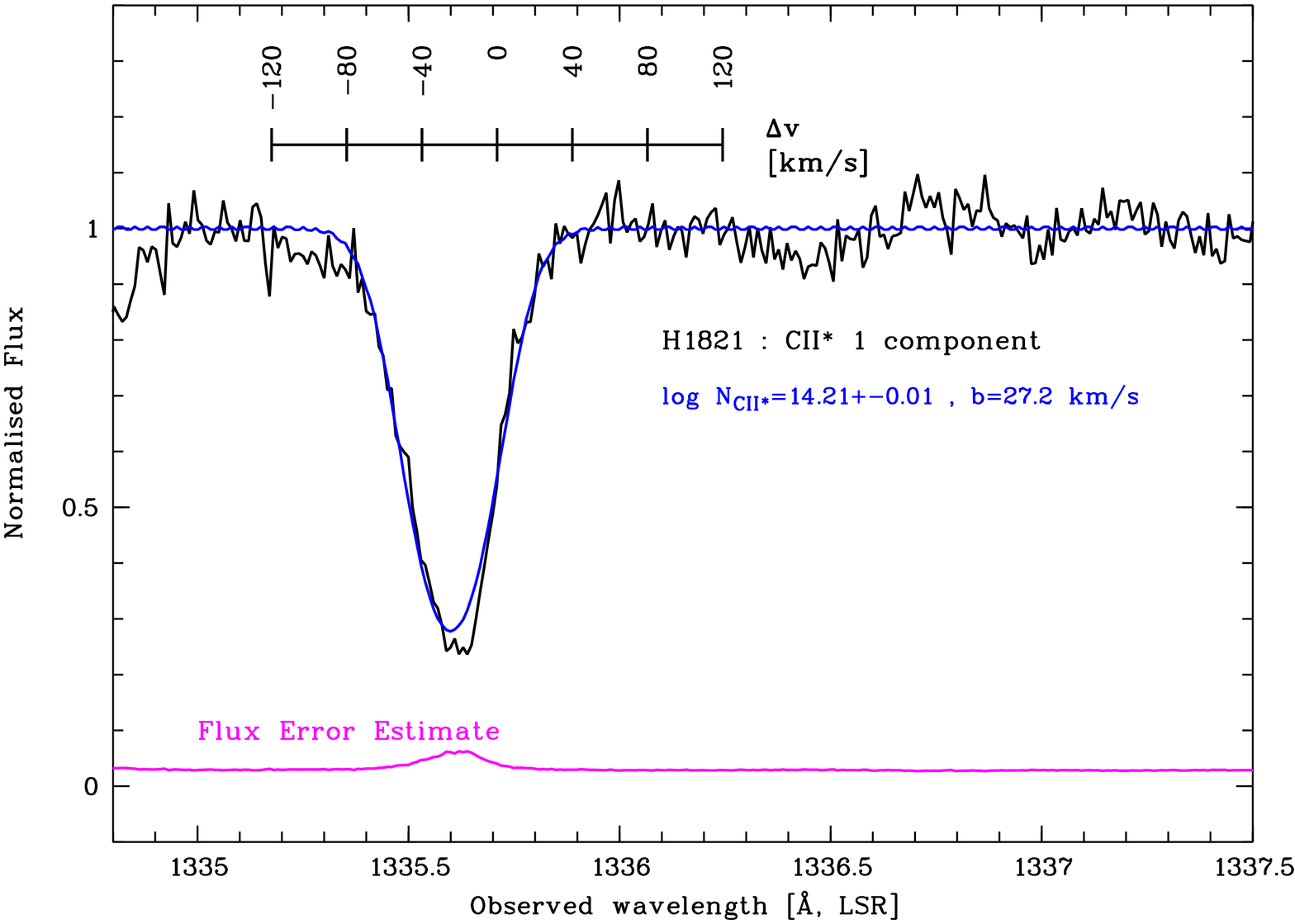}
\vspace*{-3mm}
\hspace*{-12mm}
\includegraphics[height=165mm, angle=-90.0]{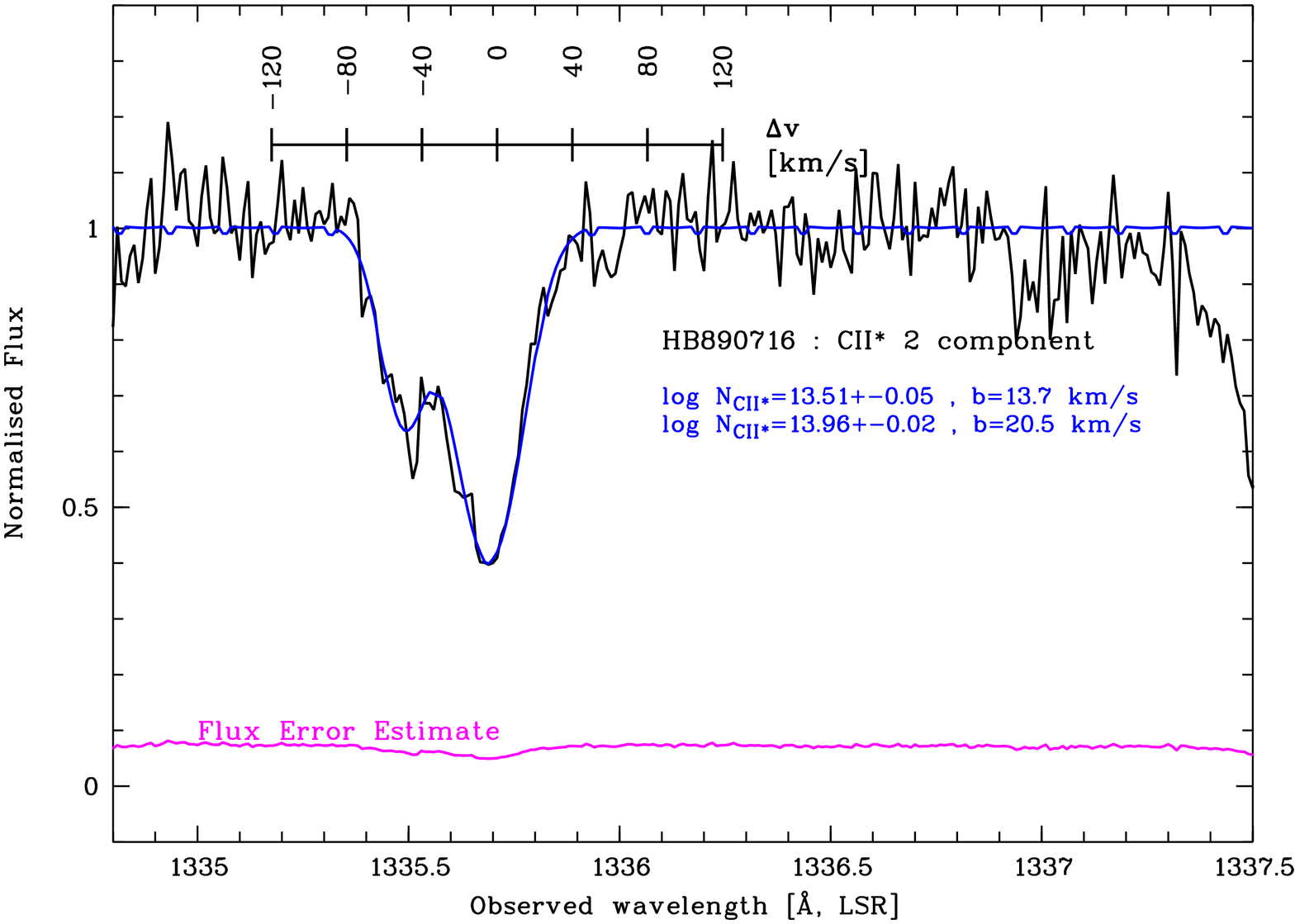}
\caption{Voigt profile fitting (continued). Top: H~1821+643 (v$_{\rm LSR} \approx -24.2$ km~s$^{-1}$); Bottom: HB89~0716+714 (v$_{\rm LSR} \approx -48.9$ and $-4.0$ km~s$^{-1}$).}
\end{center}
\end{figure*}

\setcounter{figure}{5}
\begin{figure*}
\begin{center}
\vspace*{-4mm}
\hspace*{-12mm}
\includegraphics[height=165mm, angle=-90.0]{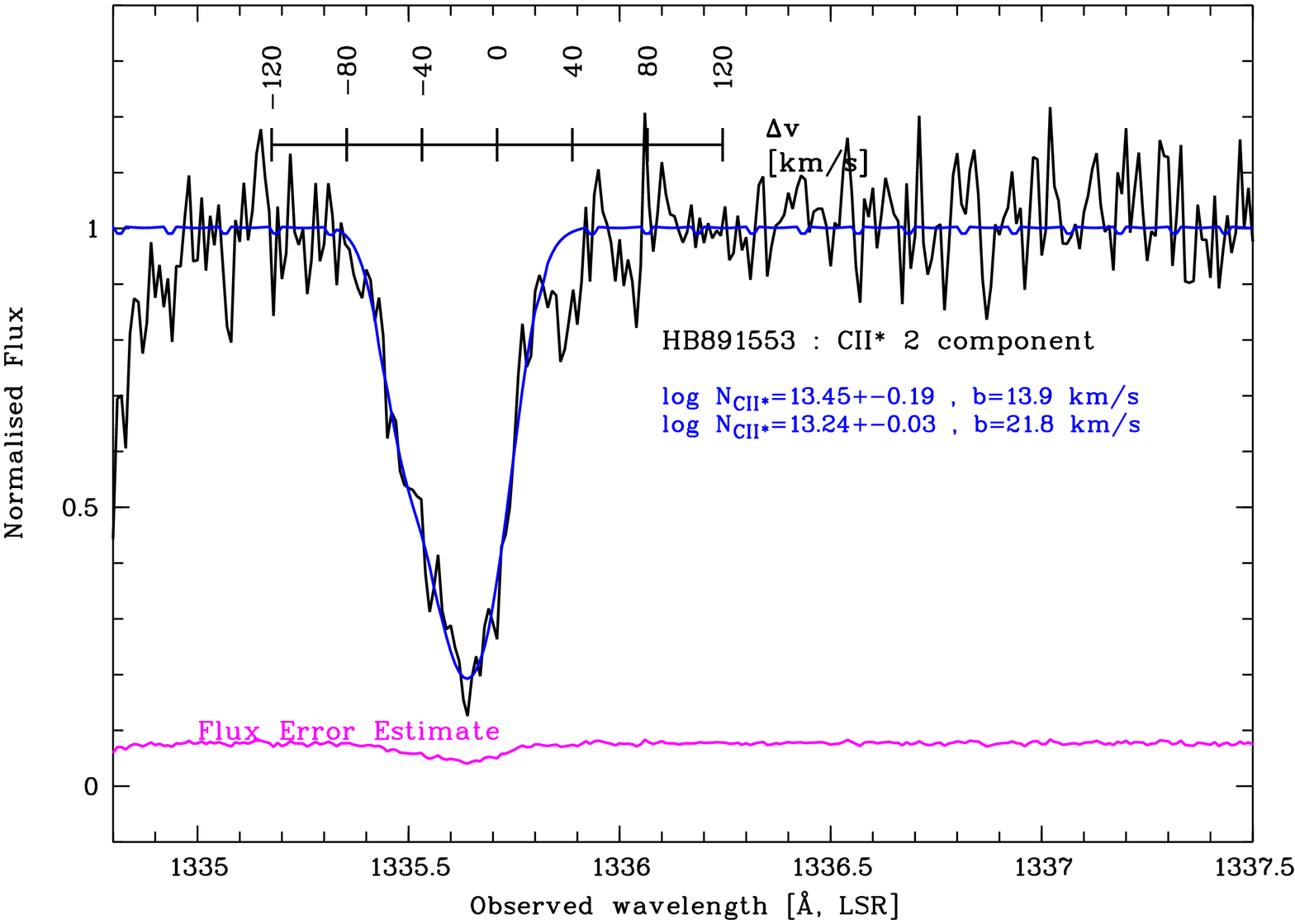}
\vspace*{-3mm}
\hspace*{-12mm}
\includegraphics[height=165mm, angle=-90.0]{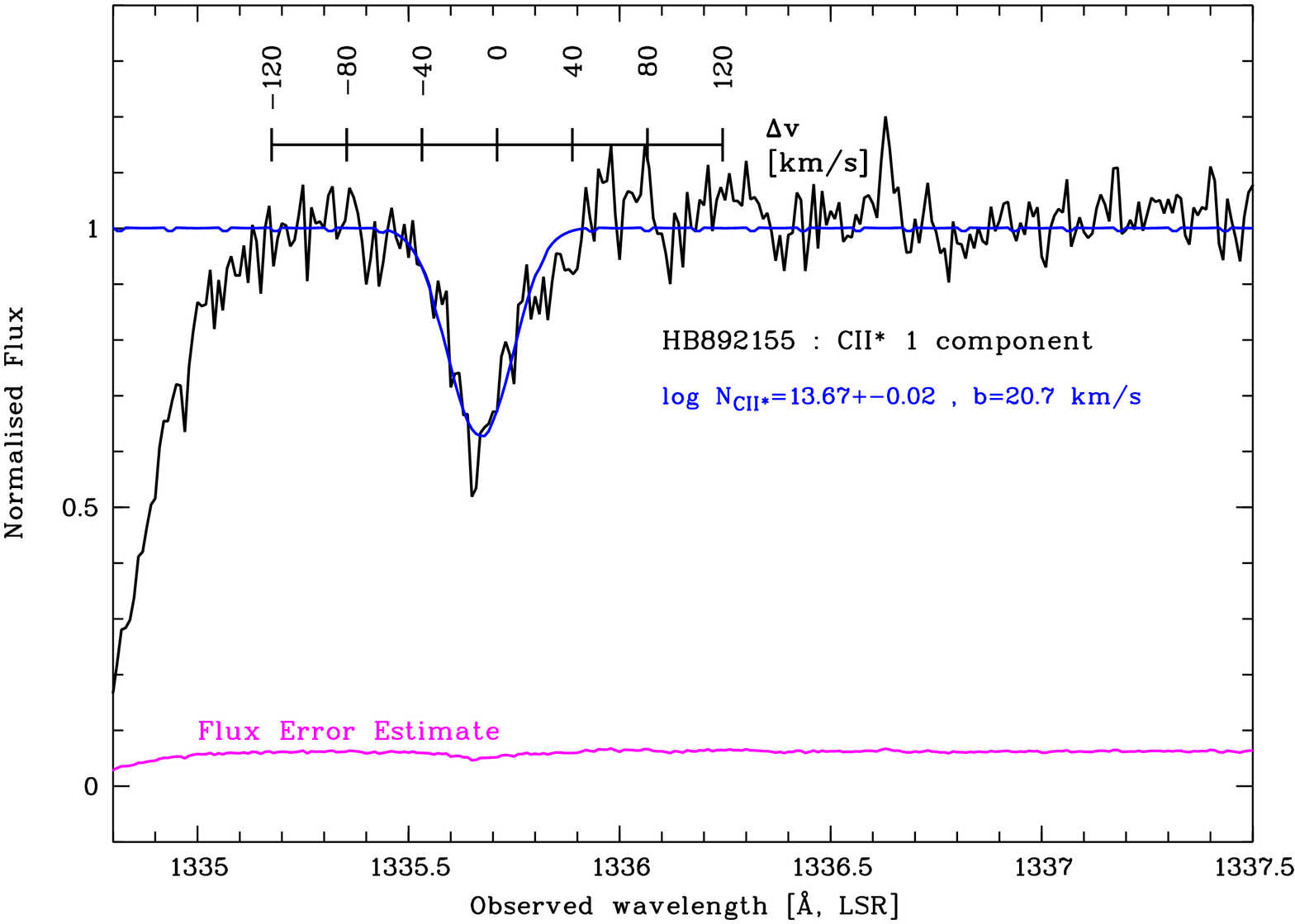}
\caption{Voigt profile fitting (continued).Top: HB89~1553+113 (v$_{\rm LSR} \approx -48.9$ and $-15.2$ km~s$^{-1}$); Bottom: (v$_{\rm LSR} \approx -8.4$ km~s$^{-1}$).}
\end{center}
\end{figure*}

\setcounter{figure}{5}
\begin{figure*}
\begin{center}
\vspace*{-4mm}
\hspace*{-12mm}
\includegraphics[height=150mm, angle=-90.0]{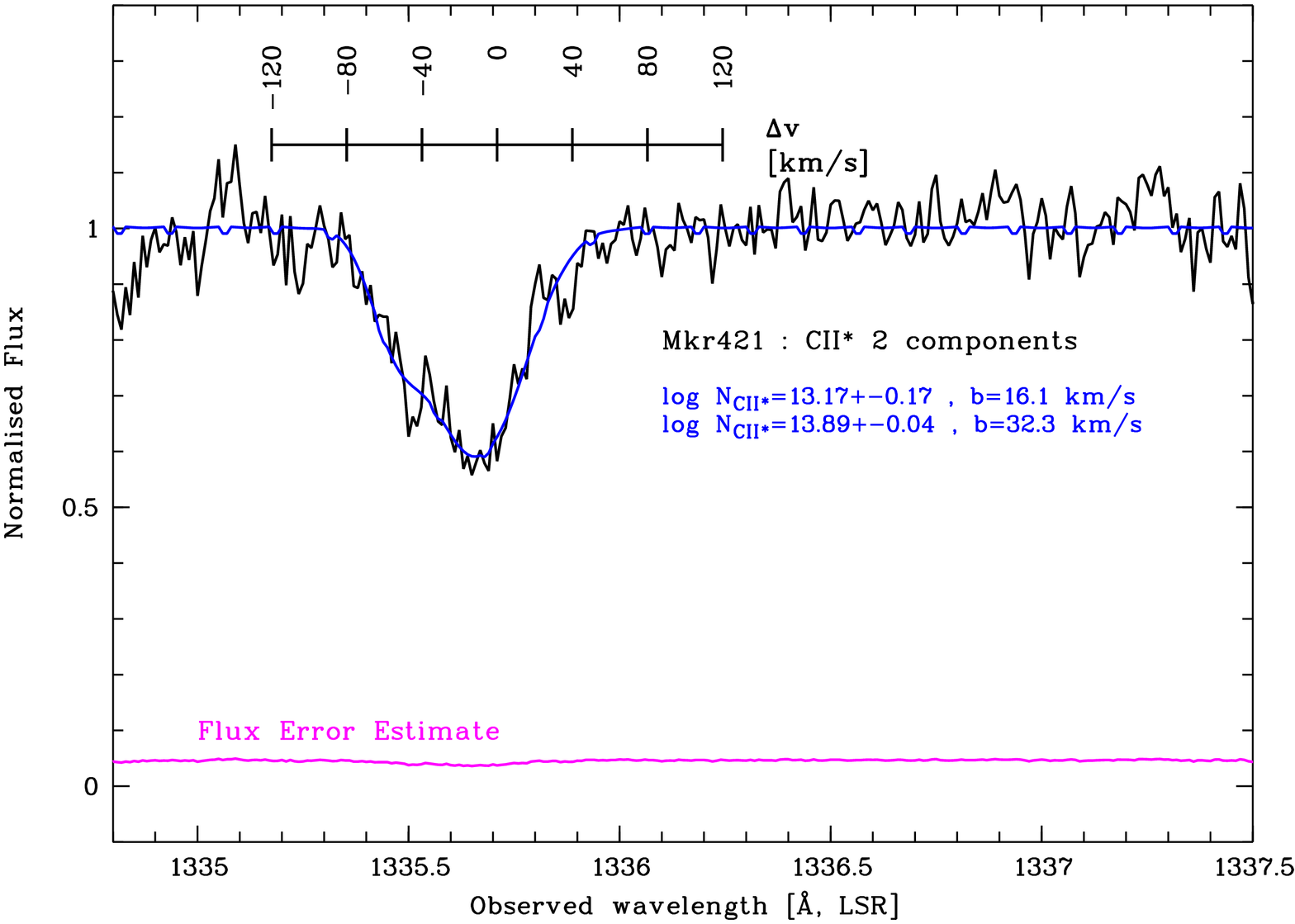}
\vspace*{-3mm}
\hspace*{-12mm}
\includegraphics[height=150mm, angle=-90.0]{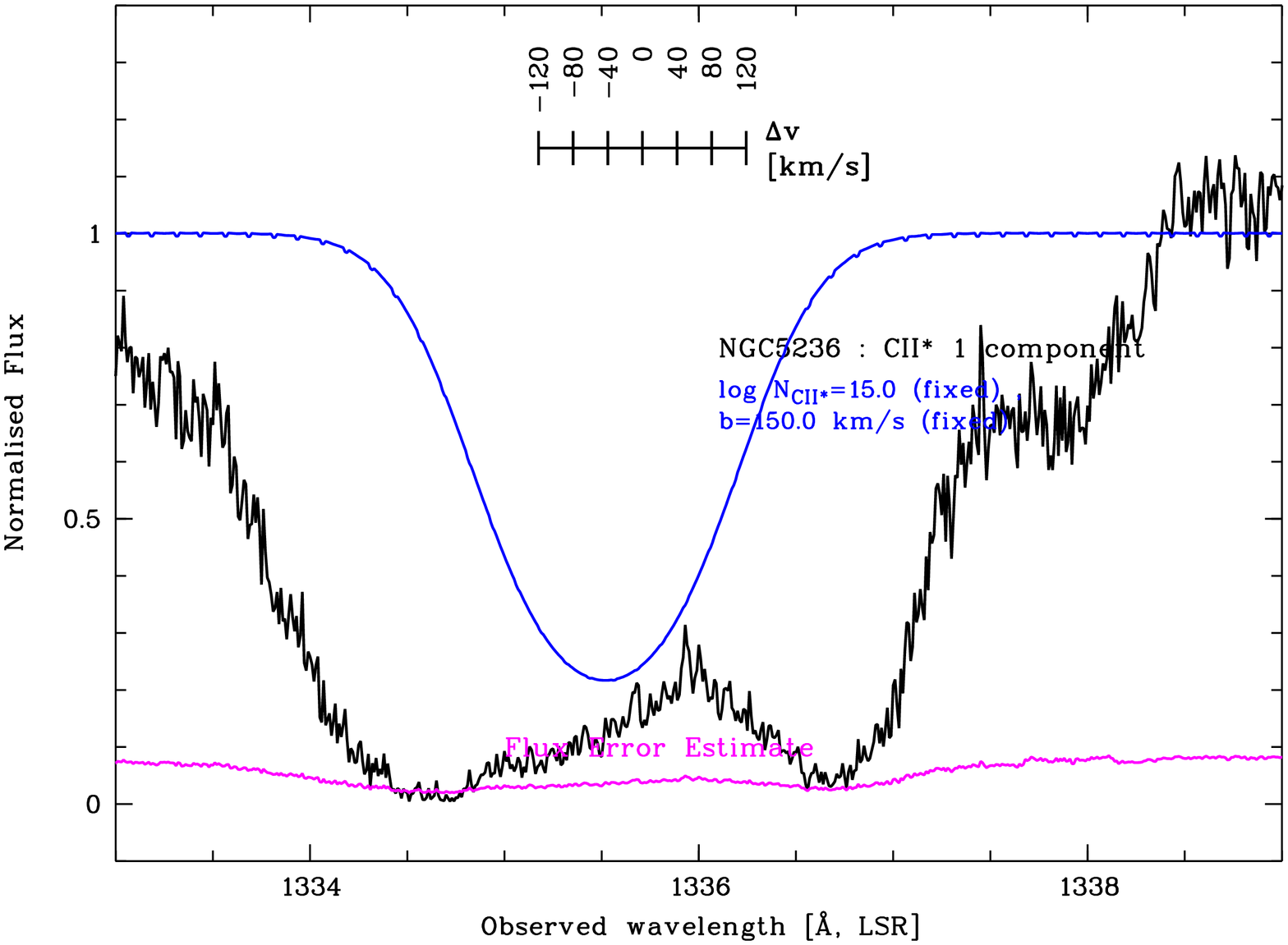}
\caption{Voigt profile fitting (continued). Top: Mrk~421 (v$_{\rm LSR} \approx -53.4$ and $-10.7$ km~s$^{-1}$); Bottom: NGC~5236 (v$_{\rm LSR} \approx -42.1$ km~s$^{-1}$); For NGC~5236, please note the presence of the very strong Galactic ISM component along with the C~{\sc ii} 1334 and 1335 \AA~ feature of the QSO host, redshifted slightly to 1336.7 \AA. In order to obtain a reasonable fit for the Galactic C~{\sc ii}$^*$ component, apart from including these transitions, we had to fix the parameters of the C~{\sc ii}$^*$ transition line to some reasonable value. Hence, the values for the column density, b parameter and velocity offset in this case should be taken with caution.}
\end{center}
\end{figure*}

\setcounter{figure}{5}
\begin{figure*}
\begin{center}
\vspace*{-4mm}
\hspace*{-12mm}
\includegraphics[height=165mm, angle=-90.0]{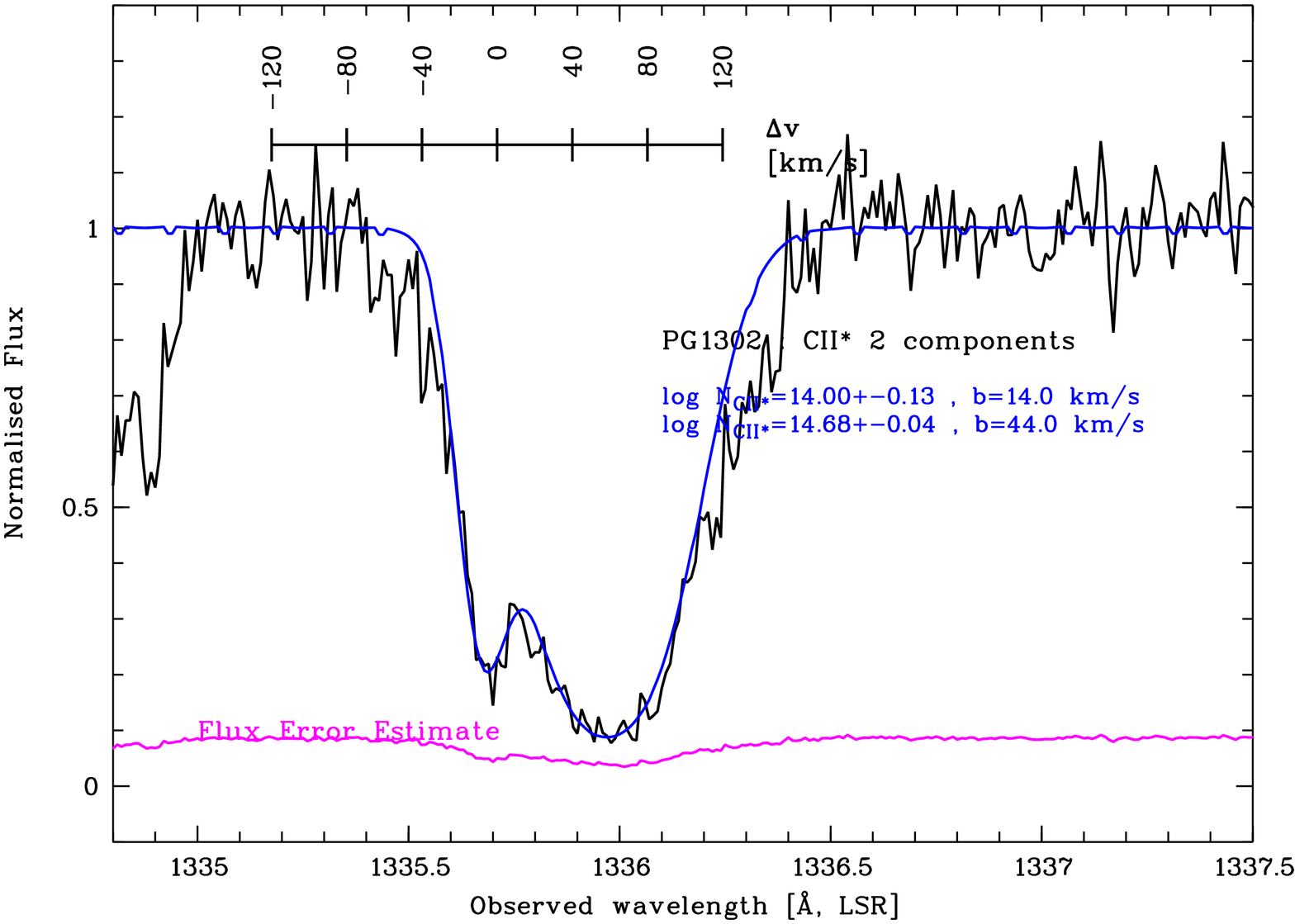}
\vspace*{-3mm}
\hspace*{-12mm}
\includegraphics[height=165mm, angle=-90.0]{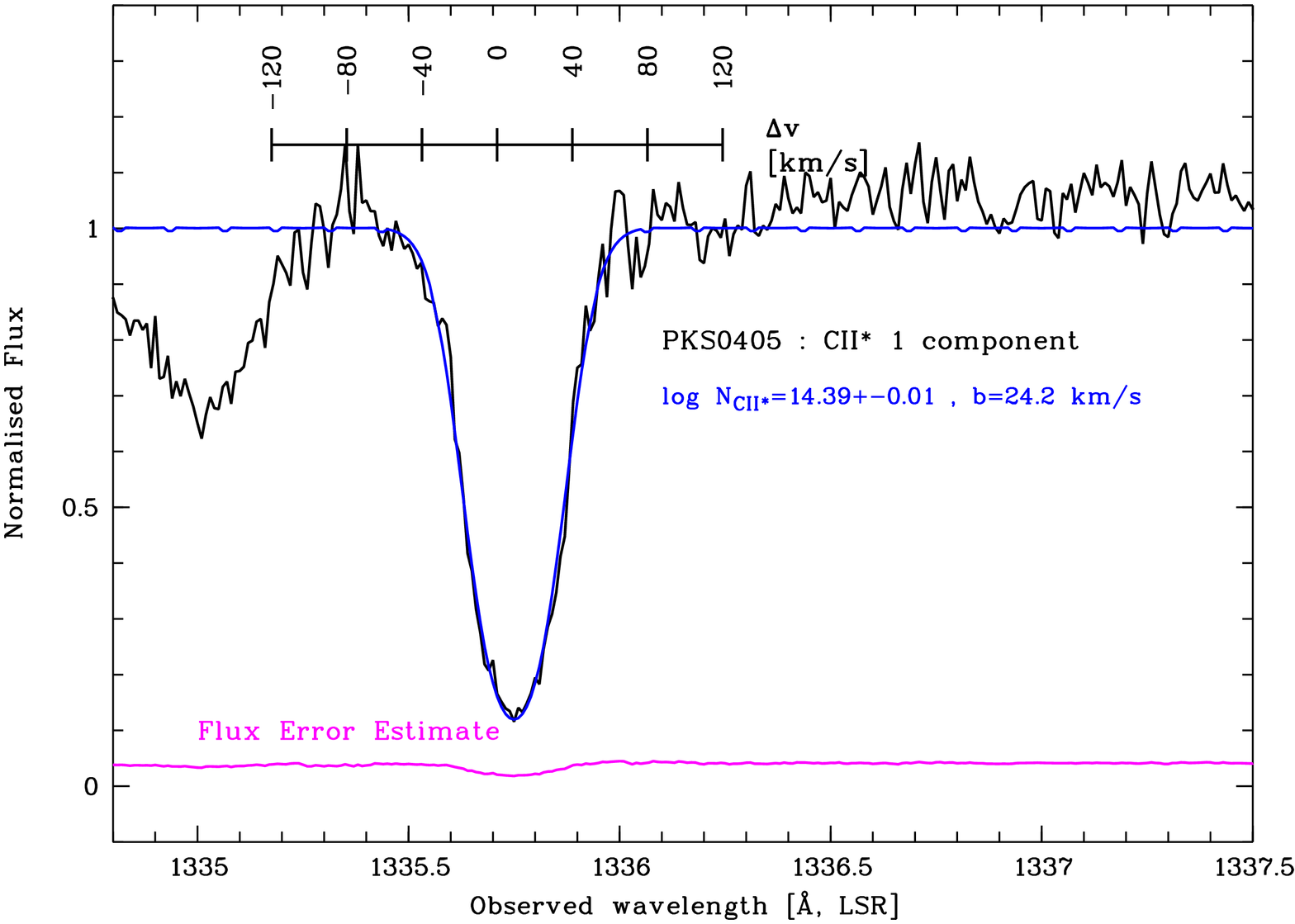}
\caption{Voigt profile fitting (continued). Top: PG~1302-102 (v$_{\rm LSR} \approx -6.2$ and $+58.9$ km~s$^{-1}$); Bottom: PKS~0405-12 (v$_{\rm LSR} \approx 9.5$ km~s$^{-1}$).}
\end{center}
\end{figure*}

\begin{figure*}
\begin{center}
\includegraphics[scale=0.225, angle=-90.0]{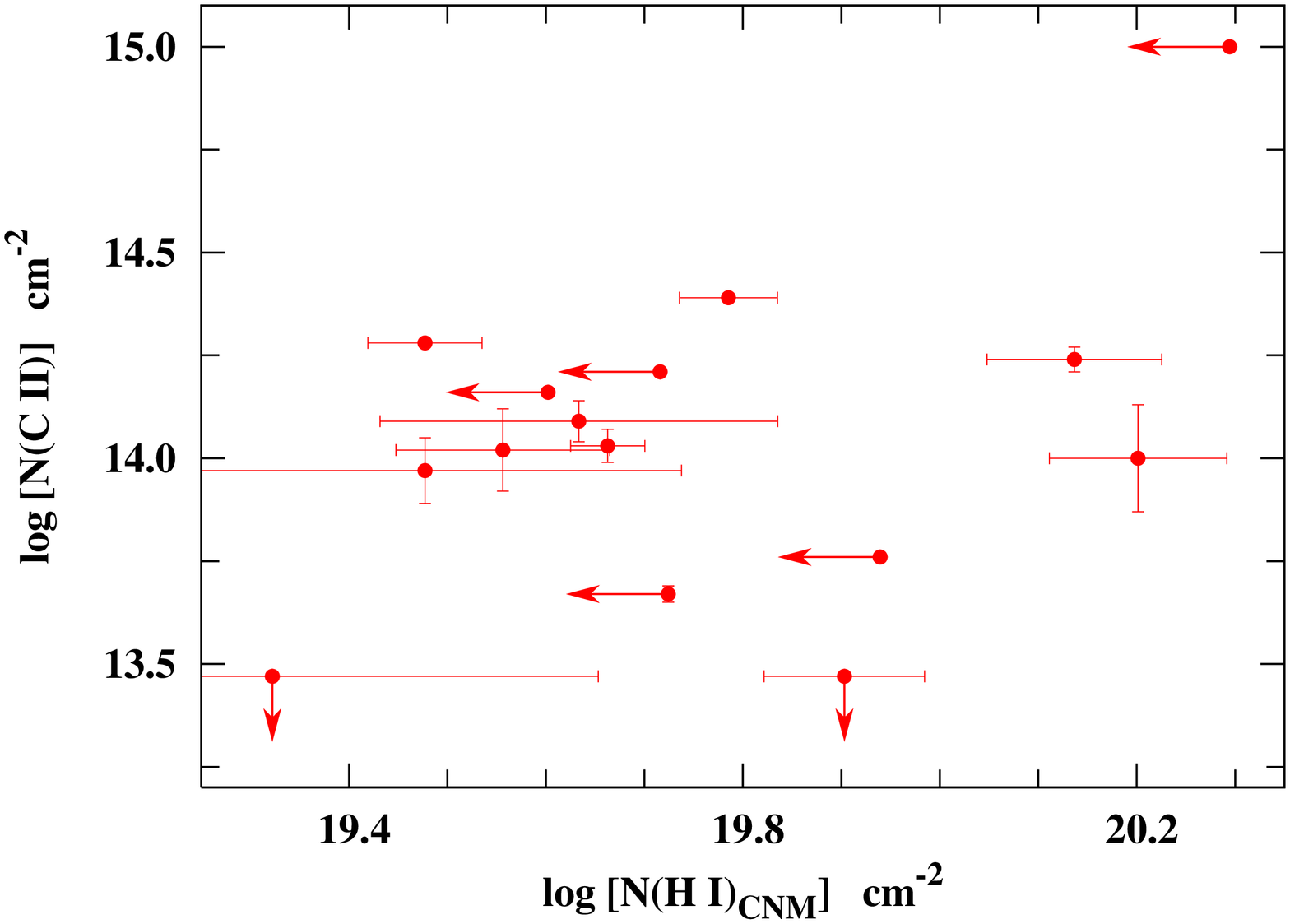}\includegraphics[scale=0.225, angle=-90.0]{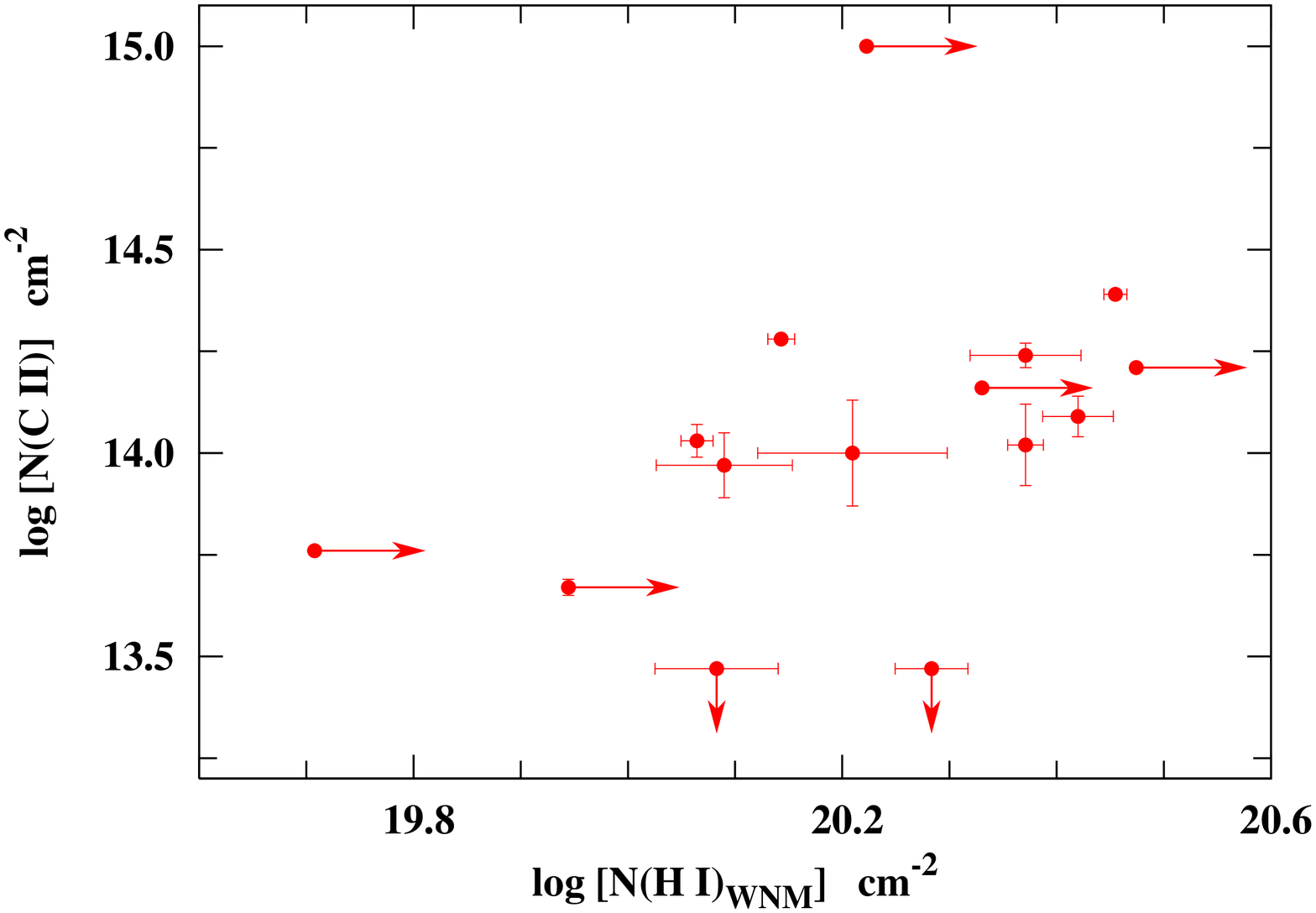}\includegraphics[scale=0.225, angle=-90.0]{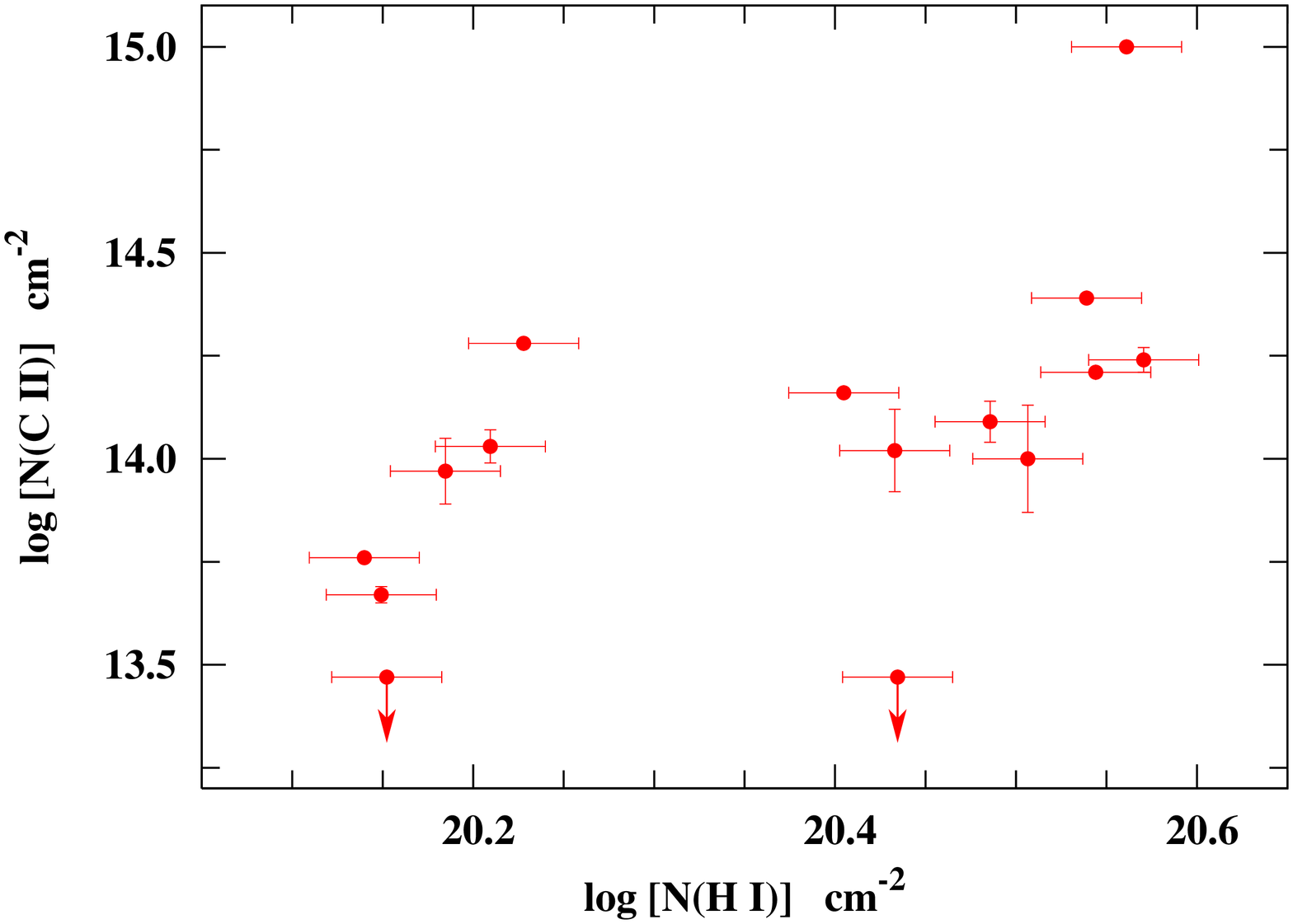}
\caption{\label{fig:fig6} Total C~{\sc ii}$^*$ column density along the lines of sight with respect to cold, warm and total H~{\sc i} column density (left, middle and right panel, respectively).}
\end{center}
\end{figure*}

\subsection{H~{\sc i} and C~{\sc ii} Column density correlations}

\begin{figure}
\begin{center}
\includegraphics[scale=0.33, angle=-90.0]{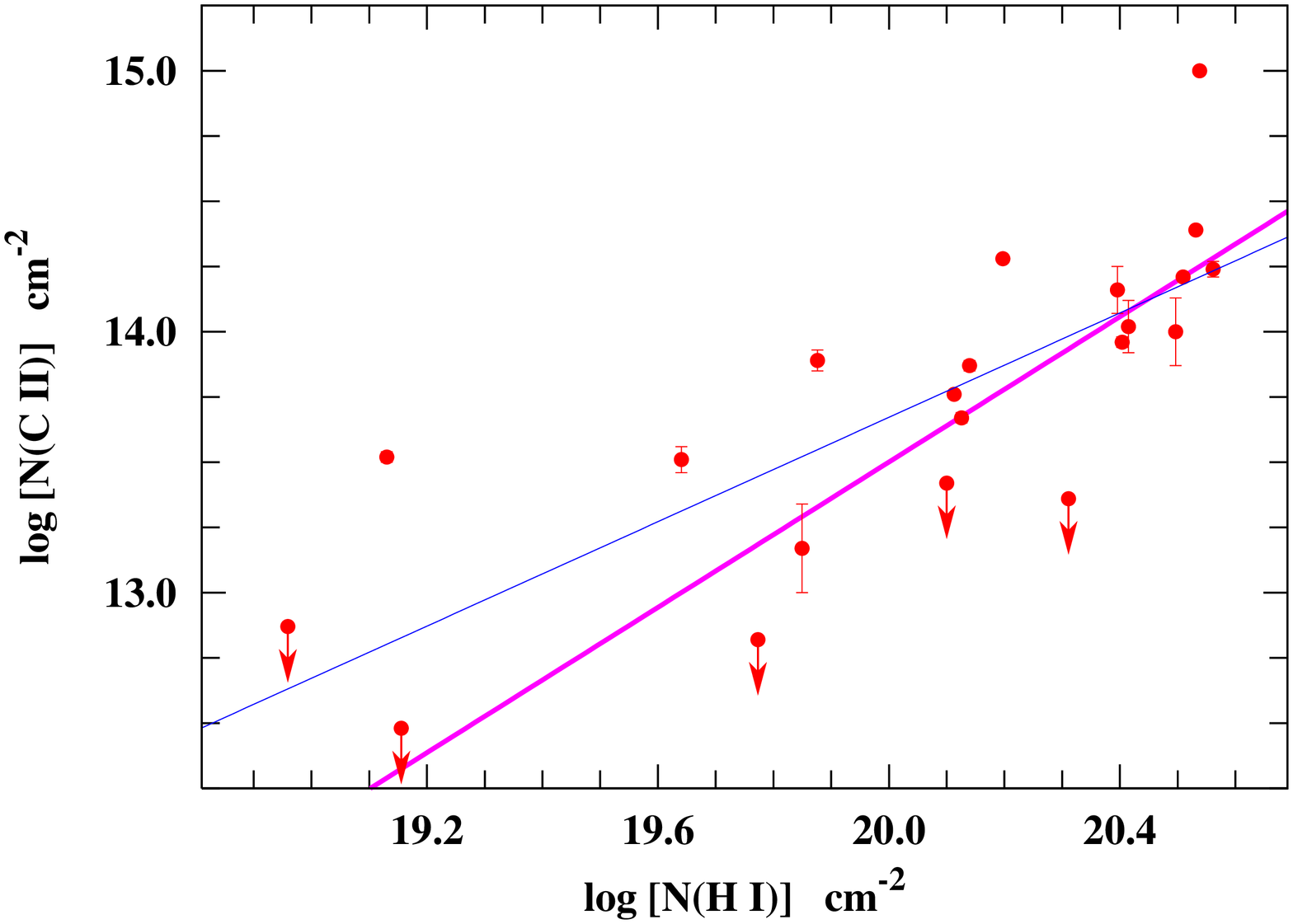}
\caption{\label{fig:fig7} Correlation between N(C~{\sc ii}$^*$) and N(H~{\sc i}) for distinct velocity components along the lines of sight. The thick line is the Akritas-Thiel-Sen regression line and the thin line represents a constant abundance of n(C~{\sc ii}$^*$)/n(H~{\sc i}) $\approx$ N(C~{\sc ii}$^*$)/N(H~{\sc i}) $= 4.7\times10^{-7}$.}
\end{center}
\end{figure}

The H~{\sc i} spin temperature measurements shows the presence of $\sim 
10-30\%$ cold gas along these lines of sight at the same velocities as the 
C~{\sc ii}$^*$ absorption components. This, however, does not mean that 
C~{\sc ii}$^*$ coexists with only cold gas. One way to investigate if 
C~{\sc ii} cooling takes place preferably in the cold or warm phase is to check 
how the derived total C~{\sc ii}$^*$ column density depends on the cold and/or 
warm H~{\sc i} column density for these lines of sight. If the C~{\sc ii}$^*$ 
absorption arises dominantly in cold or warm phase, then one will expect a 
tight correlation between N(C~{\sc ii}$^*$) and N(H~{\sc i})$_{\rm CNM}$ or 
N(H~{\sc i})$_{\rm WNM}$, respectively. Alternatively, if the C~{\sc ii}$^*$ 
abundance is similar in the cold and warm phases, there will be stronger 
correlation between N(C~{\sc ii}$^*$) and the total N(H~{\sc i}).

The three panels in Figure~\ref{fig:fig6} show the total C~{\sc ii}$^*$ column 
density along the lines of sight of this sample with respect to the cold, warm 
and the total H~{\sc i} column density (left, middle and right panel, 
respectively). We do not see any obvious strong correlation in these plots. 
This is expected when a significant fraction of C~{\sc ii}$^*$ is in the WIM 
phase, as suggested by, e.g. \citet{le04}. A careful and quantitative 
statistical analysis, however, shows that the correlation is relatively 
stronger between N(C~{\sc ii}$^*$) and the total N(H~{\sc i}) compared to the 
other two. The Kendall $\tau$ coefficient is $0.56$ for the correlation with 
the total N(H~{\sc i}) with a two-sided p-value of $0.0087$. The $\tau$ 
coefficients for N(H~{\sc i})$_{\rm CNM}$ and N(H~{\sc i})$_{\rm WNM}$ are 
$0.11$ (p $= 0.80$) and $0.40$ (p $= 0.21$), respectively. We also computed the 
Spearman's rank correlation coefficients. The correlation coefficient in this 
case is $\rho = 0.74$ for the total N(H~{\sc i}), but $0.12$ and $0.51$ for 
the cold and warm gas column density, respectively. Clearly, the correlation is 
relatively tighter for the total N(H~{\sc i}). The above values of $\tau$, p 
and $\rho$ are computed using lines of sight excluding the non-detections. 
Including the non-detections (i.e. treating the $3\sigma$ limits as ``measured 
values''), and/or excluding NGC~5236 (for which C~{\sc ii}$^*$ column density 
measurement was problematic) in the analysis, do 
not change the results significantly. Even then, N(C~{\sc ii}$^*$) shows a 
relatively tighter correlation with the total N(H~{\sc i}). This correlation, 
though not visually obvious, indicates that the C~{\sc ii}$^*$ abundance is 
not significantly different in the cold and warm phases.

As a simple consistency check, we considered the correlation between 
N(C~{\sc ii}$^*$) and N(H~{\sc i}) for individual velocity components. The 
result is shown in Figure~\ref{fig:fig7}. Here also the data are consistent 
with a significant correlation between the C~{\sc ii}$^*$ column density and 
the total H~{\sc i} column density (Kendall $\tau = 0.64$ with p $= 
5.383\times10^{-5}$ using generalized Kendall's $\tau$ test to include the 
non-detections\footnote{This is done using the Nondetects And Data Analysis 
(`NADA') package \citep{lee13} in the R statistical software environment 
\citep{rct15} available from the Comprehensive R Archive Network (CRAN) 
site.}), and with no obvious separation of CNM/WNM components. 
This again suggests a similar abundance of C~{\sc ii}$^*$ in the cold and warm 
phases. The thin blue line in Figure~\ref{fig:fig7} shows a constant value of 
N(C~{\sc ii}$^*$)/N(H~{\sc i}) = $4.7\times10^{-7}$ corresponding to the median 
abundance of the sample, and the thick magenta line is the best fit 
Akritas-Thiel-Sen regression line calculated in NADA by consistently including 
the non-detections also. For the range of column densities of our interest, as 
shown in Figure~\ref{fig:fig7}, these two do not differ much.

As an extension to this, we have also tried a multivariate linear regression 
analysis to separate out CNM, WNM and WIM contribution to the total observed 
C~{\sc ii}$^*$ column density. We have used the measured CNM and WNM column 
densities from this study, and the H$\alpha$ intensities from the Wisconsin 
H-Alpha Mapper (WHAM) Survey \citep{haf03} as a proxy for WIM column densities. 
This analysis suggests a relation
\begin{eqnarray}
\frac{N(C II^*)}{10^{13}~cm^{-2}} &=& (1.43 \pm 0.84)\times \frac{N(CNM)}{10^{19}~cm^{-2}}\nonumber\\
& + & (0.49 \pm 0.12)\times \frac{N(WNM)}{10^{19}~cm^{-2}}\nonumber\\
& + & (0.02 \pm 0.49)\times \frac{I(H\alpha)}{1~Rayleigh}
\end{eqnarray}
indicating a very weak dependence on N(WIM)in contrast to the \citet{le04} 
assertion that most of the C~{\sc ii}$^*$ is in the WIM phase. The coefficient 
for N(CNM) also has a large uncertainty (effectively consistent with zero at 
$< 2\sigma$ level), indicating strongest ($> 4\sigma$) dependence of 
N(C~{\sc ii}$^*$) on N(WNM). However, if we drop the dependence on 
I(H$\alpha$), the regression analysis results in
\begin{eqnarray}
\frac{N(C II^*)}{10^{13}~cm^{-2}} &=& (0.59 \pm 0.12)\times \frac{N(CNM)}{10^{19}~cm^{-2}}\nonumber\\
& + & (0.10 \pm 0.31)\times \frac{N(WNM)}{10^{19}~cm^{-2}}
\end{eqnarray}
implying most of the C~{\sc ii}$^*$ to be existing in the CNM. In this case, 
the coefficient for N(WNM) has a large uncertainty and a low statistical 
significance. The inconsistent and contradictory result of this analysis may be 
due to the small number of lines of sight in our study, and it should be 
carried out for a larger sample to properly separate out C~{\sc ii}$^*$ column 
densities in different ISM phases. As the multivariate regression analysis 
remains inconclusive in this case, based on the observed correlations mentioned 
earlier, the abundance of C~{\sc ii}$^*$ is assumed to be same in the cold and 
warm neutral medium for all further analysis.

\subsection{The estimated star formation rate}

We have also derived, by a method similar to that used for the DLAs, the SFR 
per unit area from the observed quantities for our lines of sight. An accurate 
estimation of the interstellar radiation field, and therefore the SFR from the 
observed C~{\sc ii}$^*$ and H~{\sc i} column density, involves computation of 
thermal and ionization equilibrium conditions in the presence of all of the 
relevant heating and cooling mechanisms. This is possible only via detailed 
numerical simulations, which is beyond the scope of the present work. Rather 
we used some simplifying assumptions, and results from existing simulations 
\citep[e.g.][]{wo95,wo03} to estimate an approximate value of the SFR. So, the 
following result should be considered as an order of magnitude consistency 
check.

\begin{table*}
 \caption{Summery of the results}
\begin{center}
 \begin{tabular}{lccccc}
 \tableline
Background & \multicolumn{2}{c}{N(H~{\sc i}) [$10^{19}$ cm$^{2}$]} & \multicolumn{3}{c}{SFR $10^{-3}$ [M$_{\odot}$ yr$^{-1}$ kpc$^{-2}$]}\\
Sources    & CNM $^{a}$ & WNM                     & CNM & WNM & two-phase \\
\tableline    
3C~71$^b$            & 3.6 $\pm$ 0.9  & 23.5       & 0.31 & ~34.2 & 14.6\\
3C~273$^b$           & 4.6 $\pm$ 0.4  & 11.6       & 0.53 & ~58.6 & 13.2\\
3C~351 $^c$          & 3.0 $\pm$ 0.4  & 13.9       & 0.90 & ~99.9 & 34.5\\
H~1821$+$643 $^c$    & $\leq$5.2      & $\geq$29.8 & 0.37 & ~41.0 & 16.2\\
HB89~0716$+$714 $^b$ & 4.3 $\pm$ 2.0  & 26.3       & 0.32 & ~35.6 & 14.6\\
HB89~1553$+$113 $^b$ & 13.7 $\pm$ 2.8 & 23.5       & 0.37 & ~41.4 & ~6.9\\
HB89~2155$-$304 $^b$ & $\leq$5.3      & $\geq$8.8  & 0.26 & ~29.4 & ~4.8\\
Mrk~421 $^c$         & 3.0 $\pm$ 1.8  & 12.3       & 0.48 & ~54.0 & 17.2\\
Mrk~501 $^b$         & 2.1 $\pm$ 1.6  & 12.1       & 0.17 & ~18.4 & ~7.3\\
NGC~1399 $^b$        & $\leq$8.7      & $\geq$5.1  & 0.33 & ~36.9 & ~2.6\\
NGC~4151 $^c$        & $\leq$4.0      & $\geq$21.4 & 0.45 & ~50.4 & 19.1\\
NGC~5236 $^b$        & $\leq$19.7     & $\geq$16.7 & 2.18 & 243.4 & 22.9\\
PG~1100$+$772 $^b$   & 8.0 $\pm$ 1.5  & 19.2       & 0.09 & ~~9.6 & ~2.1\\
PG~1302$-$102 $^c$   & 15.9 $\pm$ 3.3 & 16.2       & 0.25 & ~27.6 & ~3.0\\
PKS~0405$-$12 $^b$   & 6.1 $\pm$ 0.7  & 28.5       & 0.56 & ~62.9 & 21.8\\
\tableline
\end{tabular}\\
\end{center}
\begin{flushleft}
$^a$ CNM column density with conservative errors considering plausible undetected CNM component over the full velocity range of H~{\sc i} emission; ~~$^b$ H~{\sc i} absorption from the VLA and $^c$ the GMRT observations
\end{flushleft}
\label{table:table2}
\end{table*}

Following \citet{pwd79}, the cooling rate due to the [C{\sc ii}] 158 $\mu$m 
fine structure transition for a given C~{\sc ii}$^*$ column density 
N(C~{\sc ii}$^*$) and a H~{\sc i} column density N(H~{\sc i}) is given by
\begin{eqnarray}
l_{c}&=&{\rm N(C~II^*)}h\nu_{ul}A_{ul}/{\rm N(H~I)} \\
&=&2.89 \times 10^{-20} {\rm N(C~II^*)}/{\rm N(H~I)}~~ {\rm ergs~s}^{-1}({\rm~H~atom})^{-1} \nonumber
\label{eqn:ccr}
\end{eqnarray}
where $h\nu_{ul}$ and $A_{ul}$ are the upper level energy and coefficient for 
spontaneous decay of the $^2P_{3/2}$ to $^2P_{1/2}$ transition. On the other 
hand, the photoelectric heating rate is a function of the FUV field 
\citep{bt94,wo95,wo03}, and given by 
\begin{equation}
\Gamma_{\rm d} = 1.3 \times 10^{-24} \epsilon G_0 ~~ {\rm ergs~s}^{-1}({\rm~H~atom})^{-1}
\label{eqn:dhr}
\end{equation}
where $\epsilon$ is the heating efficiency. The FUV field strength $G_0$, 
normalized to the local interstellar value \citep[$1.6 \times 10^{-3}$ 
ergs~cm$^{-2}$~s$^{-1}$;][]{ha68}, is proportional to the SFR per unit area. 
For an SFR of $\log_{10}\dot{\psi_*} = -2.4$ M$_{\odot}$ yr$^{-1}$ kpc$^{-2}$, 
$G_0 = 1.7$ \citep*{dr78}.

To estimate $\dot{\psi_*}$ using the total N(H~{\sc i}) from the LAB survey, 
and the CNM/WNM column density from this work, we now consider three situations 
where these lines of sight have (i) only cold gas (``CNM model''), (ii) only 
warm gas (``WNM model''), and (iii) a mix of CNM and WNM (``two-phase model'') 
with the same abundance ratio of C~{\sc ii}$^*$ to H~{\sc i} column density. 
The photoelectric heating rate from dust grains is approximately equal to the 
[C~{\sc ii}] fine structure cooling rate in the CNM. On the other hand, the 
[C~{\sc ii}] fine structure cooling rate in the WNM is about an order of 
magnitude lower than the photoelectric heating rate. On average, the 
photoelectric heating accounts for $\approx 60 - 65\%$ of the total heating 
rate. The heating efficiencies are $\epsilon \sim 0.05$ and $\sim 0.005$ for 
the cold and the warm phase, respectively \citep{wo95}.

\begin{figure}
\begin{center}
\includegraphics[scale=0.33, angle=-90.0]{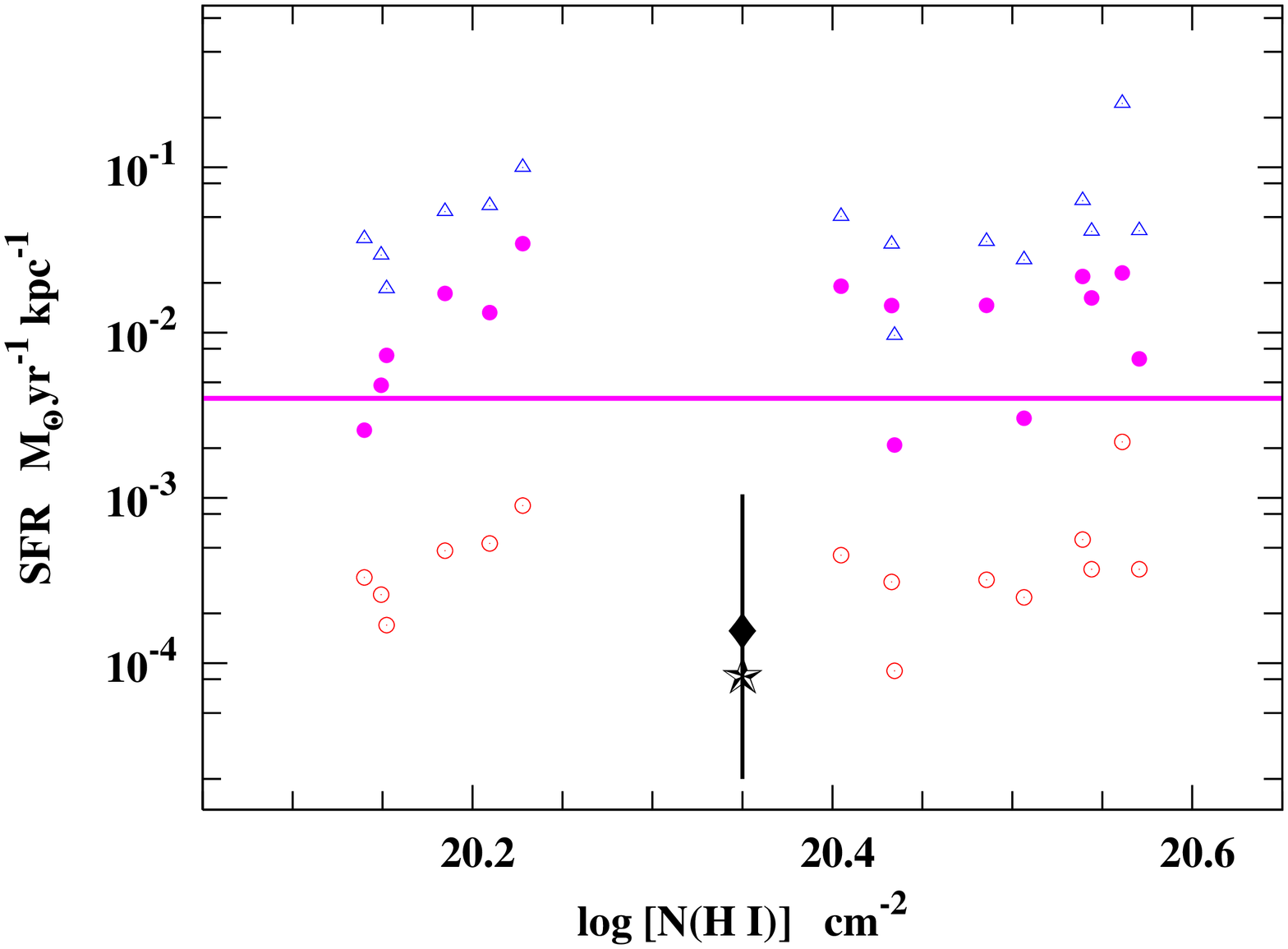}
\caption{\label{fig:fig8} Estimated SFR from observed N(C~{\sc ii}$^*$) and N(H~{\sc i}). Different symbols are for ``CNM model'' (open circle), ``WNM model'' (open triangle) and ``two-phase model'' (filled circles). The solid line marks the SFR of $\sim 4 \times 10^{-3}$ M$_{\odot}$ yr$^{-1}$ kpc$^{-2}$ for the Milky Way \citep{k98}. The vertical line with diamond and star shows the range, mean and median SFR based on the correlation reported in \citet{herr14}.}
\end{center}
\end{figure}

Putting these numbers together, and using the observed N(C~{\sc ii}$^*$) and 
N(H~{\sc i}) in the cold and warm phases, the SFR per unit area is estimated 
for our lines of sight (see Table~\ref{table:table2}). As shown in 
Figure~\ref{fig:fig8}, $\dot{\psi_*}$ is about $0.1 - 2 \times 10^{-3}$ 
M$_{\odot}$ yr$^{-1}$ kpc$^{-2}$ for the ``CNM model'', and about $1 - 24 
\times 10^{-2}$ M$_{\odot}$ yr$^{-1}$ kpc$^{-2}$ for the ``WNM model''. The 
median values are $0.37 \times 10^{-3}$ and $4.1 \times 10^{-2}$ M$_{\odot}$ 
yr$^{-1}$ kpc$^{-2}$ for the CNM and WNM model, respectively. For the 
``two-phase model'', the estimated range is about $2 - 35 \times 10^{-3}$ 
M$_{\odot}$ yr$^{-1}$ kpc$^{-2}$, with a median value of $14.6 \times 10^{-3}$ 
M$_{\odot}$ yr$^{-1}$ kpc$^{-2}$ . Clearly, the ``two-phase model'' provides a 
relatively better match with the Milky Way value of $\dot{\psi_*} \sim 4 \times 
10^{-3}$ M$_{\odot}$ yr$^{-1}$ kpc$^{-2}$ derived from other independent 
observations \citep*[e.g.][]{k98,ke12}. The assumption, that all the 
C~{\sc ii}$^*$ absorption arises in the WNM or CNM, respectively overpredicts 
or underpredicts $\dot{\psi_*}$ by more than an order of magnitude. 

There are studies \citep[e.g.][]{looz14,herr14}, based mostly on nearby 
galaxy samples, to calibrate C~{\sc ii} surface brightness and/or luminosity 
against other independent measurements of SFR. For a comparison, we have also 
used such correlation to derive SFR for our sample. This is done by first 
converting C~{\sc ii}$^*$ column density to a cooling rate (using 
equation~3), and then converting it to $\dot{\psi_*}$ using 
equation~2 from \citet{herr14}. The estimated range of $\dot{\psi_*}$, shown 
in Figure~\ref{fig:fig8}, is $\sim 0.2 - 10.5 \times 10^{-4}$ M$_{\odot}$ 
yr$^{-1}$ kpc$^{-2}$, with a median (mean) value of $0.84$ ($1.56$) $\times 
10^{-4}$ M$_{\odot}$ yr$^{-1}$ kpc$^{-2}$. Similar to that of the ``CNM 
model'', this underpredicts $\dot{\psi_*}$ by more than an order of magnitude. 
This discrepancy is not surprising because the high Galactic latitude lines of 
sight of our sample may have different physical conditions (including cold/warm 
phase fraction) than the ones used to derive the relation. Similarly, without 
considering these properties in detail, such relations will not be readily 
useful in estimating the SFR for the high redshift DLAs as well.

\section{Discussion}
\label{sec:dis}

It has been earlier established by \citet{le04} that the WIM is one of the major 
ISM components along high Galactic latitude lines of sight with C~{\sc ii}$^*$ 
absorption. The current study establishes from direct H~{\sc i} observations 
that the lines of sight certainly pass through a non-negligible fraction of 
CNM, coexisting in the same velocity range as that of the C~{\sc ii}$^*$ 
absorption. Computing the SFR, assuming the ISM to be in the warm phase 
(without taking the CNM fraction into consideration), results in a higher value 
of $\dot{\psi_*}$. However, even after including the CNM fraction in the 
calculation, the median value of $\dot{\psi_*}$ is more than a factor of three 
higher than the average SFR of the Milky Way. This mismatch is not surprising 
because, in our simplified order of magnitude calculation of the SFR, we have 
neglected the WIM component which is likely to contain a significant fraction 
of the observed C~{\sc ii}$^*$ column density \citep{le04}. Also, the derived 
values of $\dot{\psi_*}$ vary by a factor of about $20$ for different lines of 
sight. 

Earlier studies have suggested a correlation between [C~{\sc ii}] luminosity 
and the SFR both at galaxy scale \citep[e.g.][]{bose02,looz11}, as well as at 
smaller scales \citep[e.g.][]{mook11,kapa14,pine14}. But, it is also well-known 
that the scatter in this relation is fairly large 
\citep[e.g.][]{malh97,ke12,looz14,herr14}. For a sample of galaxies, 
\citet{looz11} found that the data are consistent with C~{\sc ii} primarily 
being associated with cold ISM; \citet{pine13} found only $\sim 4\%$ of the 
C~{\sc ii} to be from ionized gas in the Galactic plane. \citet{pine14}, on 
the other hand, concluded that multiple ISM components significantly contribute 
to the [C~{\sc ii}] luminosity of the Milky Way disk (but also see 
\citealt{geri14} who concluded that, in the Galactic plane, C~{\sc ii} has the 
same filling factor as that of the CNM). \citet{pine14} have suggested that 
the [C~{\sc ii}] luminosity -- SFR scaling relation is different for different 
ISM phases. This may be due to varying energetics, or different timescale of 
SFR that C~{\sc ii} is sensitive to in these various phases 
\citep[see][]{kapa14}. When averaged over galaxy scale, this gives rise to the 
observed extra-galactic scaling. Considering all these aspects, the large 
scatter in the derived $\dot{\psi_*}$ for our sample is not surprising. Also, 
instead of averaging over galactic scale, or restricting to the CNM dominated 
disk of the Galaxy, here we are rather considering high Galactic latitude 
lines of sight with larger WNM and WIM fraction. So, in principle, a different 
N(C~{\sc ii}) - SFR relation for this sample is quite possible.

There are two more possible reasons for the large scatter in estimated values 
of $\dot{\psi_*}$ for this sample. The Galactic value of $\dot{\psi_*} \sim 3.3 
- 4 \times 10^{-3}$ M$_{\odot}$ yr$^{-1}$ kpc$^{-2}$ \citep{k98,chom11} is only 
an average value over the entire extent of the disk. In reality, there is 
spatial variation of the SFR \citep[][]{ke12}. The other possibility is related 
to the basic assumption of the thermal steady state itself. Recently, 
theoretical as well as observational studies have raised doubts on the validity 
of the steady state model itself \citep[e.g.][]{kane03,hnt03a,hnt03b,audi05,roy13b,saur14,kim14}. There are evidences that much of the ``warm gas'' is in the 
so-called unstable phase with temperature lower than the WNM temperature of the 
classical model. More detailed studies are necessary to understand how this may 
affect the [C~{\sc ii}] luminosity -- SFR scaling relation.

{\it What are the possible connections of these results with the SFR for the 
DLAs}? It is indeed true that, for the DLAs, the C{\sc ii} cooling rate is not 
entirely balanced by the background radiation only, and requires contribution 
from a local radiation field which, in turn, is related to the star formation 
in these systems \citep[e.g.][]{wol03b,wol08,dut14}. Interestingly, the 
C~{\sc ii} cooling rate itself shows a bimodal distribution. This is proposed 
to be related to the mode of heating - {\it in situ} star formation for the 
`low cool' population (cooling rate $< 10^{-27}$ ergs~s$^{-1}$~H$^{-1}$), and 
the star formation in the central Lyman Break Galaxy for the `high cool' 
population \citep{wol08}. However, to estimate the SFR, the cold fraction of 
the gas along the line of sight still remains an important parameter 
(particularly for the low cool systems). As shown here, for the sample of 
these DLA-like lines of sight, the inferred SFR is not a good estimator of the 
Galactic average value of the true $\dot{\psi_*}$ without a proper 
characterization of the phases of the ISM in consideration. Hence, a more 
detailed analysis of the energetics of the individual systems 
\citep[e.g.][]{dut14}, as well as direct and unambiguous measurements of the 
temperature \citep[e.g.][]{how05,jor09,roy13c,kan14} is necessary to get a 
handle on the issue (please also see \citealt{wol05} for a comprehensive 
discussion).

\section{Conclusions}
\label{sec:con}

As the [C~{\sc ii}] fine structure transition is one of the main cooling 
mechanisms in the ISM, the observed C~{\sc ii} column density is often used, 
for local and high redshift systems (including DLAs), to infer the SFR. Here, 
we have used radio and UV data to study the Galactic ISM along a sample of high 
Galactic latitude ``DLA-like'' sightlines with C~{\sc ii}$^*$ absorption. The 
H~{\sc i} 21 cm absorption and emission spectra were used to directly constrain 
the temperature of the gas along these lines of sight. A good fraction ($\sim 
10 - 50\%$) of the neutral gas, even for these high latitude lines of sight, is 
in the cold phase. The correlation of the C~{\sc ii}$^*$ column density is 
tighter with the total H~{\sc i} column density (compared to that with only 
the cold or the warm gas column density). This suggests that C~{\sc ii} is 
coexisting with both the CNM and the WNM. The derived SFR values, with the 
assumption that the [C~{\sc ii}] fine structure cooling is important only in 
either the cold {\it or} the warm phase, do not match with the known value of 
the Galactic SFR. We conclude that C~{\sc ii} may not be a reliable tracer of 
star formation without adequate constraints on the temperature of the gas where 
the cooling is happening. More such multiwavelength studies of the Milky Way 
and other nearby systems, probing a wide range of ISM conditions, will be 
useful to better understand the connection between C~{\sc ii} and star 
formation for the DLAs.

\acknowledgments

This work is dedicated to the memory of A.~M.~Wolfe, who deceased on 2014 February 17, at the age of 74. 

We thank the anonymous reviewer for useful comments and suggestions that helped us improving the quality of this manuscript significantly.  We are grateful to Dr. Eric Feigelson for his valuable inputs to improve the statistical analysis. We are also grateful to Jayaram N. Chengalur, Raghunathan Srianand, Eric Greisen, Nissim Kanekar and Rajaram Nityananda for useful discussions, and to Andreas Brunthaler for his comments on an earlier version of this paper. We thank the Center for Astrophysics and Space Sciences, University of California, San Diego (CASS/UCSD) for prompt response and for permitting us to list late Prof.~A.~M.~Wolfe as a coauthor. NR acknowledges support from MPIfR and the Alexander von Humboldt Foundation during his stay at MPIfR where part of this research was carried out. This research has made use of NASA's Astrophysics Data System. We thank the staff of the GMRT and the VLA who have made these observations possible. The GMRT is run by the National Centre for Radio Astrophysics of the Tata Institute of Fundamental Research. The National Radio Astronomy Observatory is a facility of the National Science Foundation operated under cooperative agreement by Associated Universities, Inc. The Wisconsin H-Alpha Mapper is funded by the National Science Foundation. Some of the data used in this paper were obtained from the Leiden/Argentina/Bonn Galactic H~{\sc i} survey. The results reported here are also based on observations made with the NASA/ESA Hubble Space Telescope, obtained from the data archive at the Space Telescope Science Institute. STScI is operated by the Association of Universities for Research in Astronomy, Inc. under NASA contract NAS 5-26555.

\end{document}